\documentclass[a4paper,11pt]{article}
\usepackage{amssymb}
\usepackage{jcappub} 

\usepackage{doi}
\usepackage{lineno}
\usepackage{textcomp} 
\usepackage{easyReview}
\usepackage{overpic}

\title{\boldmath Blind mitigation of foreground-induced biases on primordial $B$ modes for ground-based CMB experiments}

\def\reff@jnl#1{{\rm#1\/}}

\def\aj{\reff@jnl{AJ}}                  
\def\araa{\reff@jnl{ARA\&A}}            
\def\apj{\reff@jnl{ApJ}}                
\def\apjl{\reff@jnl{ApJ}}               
\def\apjs{\reff@jnl{ApJS}}              
\def\ao{\reff@jnl{Appl.Optics}}         
\def\apss{\reff@jnl{Ap\&SS}}            
\def\aap{\reff@jnl{A\&A}}               
\def\aapr{\reff@jnl{A\&A~Rev.}}         
\def\aaps{\reff@jnl{A\&AS}}             
\def\azh{\reff@jnl{AZh}}                        
\def\baas{\reff@jnl{BAAS}}              
\def\jcap{\reff@jnl{JCAP}}              
\def\jrasc{\reff@jnl{JRASC}}            
\def\memras{\reff@jnl{MmRAS}}           
\def\mnras{\reff@jnl{MNRAS}}            
\def\pra{\reff@jnl{Phys.Rev.A}}         
\def\prb{\reff@jnl{Phys.Rev.B}}         
\def\prc{\reff@jnl{Phys.Rev.C}}         
\def\prd{\reff@jnl{Phys.Rev.D}}         
\def\prl{\reff@jnl{Phys.Rev.Lett}}      
\def\physrep{\reff@jnl{Phys.Rep.}}      
\def\pasp{\reff@jnl{PASP}}              
\def\pasj{\reff@jnl{PASJ}}              
\def\qjras{\reff@jnl{QJRAS}}            
\def\skytel{\reff@jnl{S\&T}}            
\def\solphys{\reff@jnl{Solar~Phys.}}    
\def\sovast{\reff@jnl{Soviet~Ast.}}     
 \def\ssr{\reff@jnl{Space~Sci.Rev.}}    
\def\zap{\reff@jnl{ZAp}}                
\def\nat{\reff@jnl{Nature}}  

\author[1,2]{A. Mustafa}
\author[1, 3, 4]{A. Carones}
\author[1, 3, 4]{N. Krachmalnicoff}
\author[5,6]{M. Migliaccio}
\author[1,2,3, 4]{C. Baccigalupi}
\affiliation[1]{International School for Advanced Studies (SISSA), Via Bonomea 265, 34136, Trieste, Italy}
\affiliation[2]{Università di Trento, Dipartimento di Fisica, Via Sommarive 14, 38123, Trento, Italy}
\affiliation[3]{INFN Sezione di Trieste, via Valerio 2, 34127 Trieste, Italy}
\affiliation[4]{IFPU, Via Beirut, 2, 34151 Grignano, Trieste, Italy}
\affiliation[5]{Dipartimento di Fisica, Università di Roma “Tor Vergata”, via della Ricerca Scientifica 1, I-00133, Roma, Italy}
\affiliation[6]{INFN Sezione di Roma2, Università di Roma “Tor Vergata”, via della Ricerca Scientifica, 1, 00133,
Roma, Italy}

\emailAdd{amustafa@sissa.it}

\abstract {Observations of the Cosmic Microwave Background (CMB) B-mode polarisation provide a unique probe of inflationary physics. Extracting a reliable constraint on the tensor-to-scalar ratio $r$ nonetheless demands stringent suppression of diffuse Galactic foregrounds, whose residuals can bias the inferred signal.
This work introduces and evaluates two extensions of the Needlet Internal Linear Combination (NILC) framework aimed at reducing foreground-induced biases on $r$. The first extension implements the deprojection of selected foreground moments directly within the component-separation step. The second performs a likelihood-level marginalisation over residual foreground power using a data-driven template. Using Simons Observatory Small Aperture Telescope (SO-SAT) - like simulations, we show that both methods effectively control residual contamination, yielding unbiased estimates of $r$ and a consistent reconstruction of the lensing B-mode amplitude.
These results indicate that enhanced foreground-mitigation strategies will be useful for next-generation CMB polarisation analyses seeking a robust detection of primordial B-modes.
}

\begin{document}
\maketitle
\flushbottom

\section{Introduction}
\label{sec:intro}
One of the key pillars of the standard cosmological model is the occurrence of a primordial phase of exponential expansion of the Universe, known as \emph{cosmic inflation}, which took place immediately after the Big Bang \cite{Guth1981,Linde1982}. This process stretched quantum vacuum fluctuations to cosmological scales, seeding the primordial density perturbations that later evolved into the large-scale structures observed today. In addition to these scalar perturbations, inflation is also expected to have generated a stochastic background of tensor fluctuations in the metric i.e, primordial gravitational waves, which would have left a distinctive imprint in the polarisation field of the Cosmic Microwave Background (CMB) at large angular scales (low multipoles, $\ell \lesssim 100$) \cite{Seljak1996, Kamionkowski_1998}. Specifically, during the epoch of recombination, primordial gravitational waves induce quadrupolar temperature anisotropies in the photon–baryon plasma, generating a distinct curl-like ($B$-mode) polarisation pattern on degree angular scales. This produces a characteristic peak in the angular power spectrum at $\ell \sim 80$, commonly referred to as the \emph{recombination bump}. Later, during the epoch of reionisation, scattering of CMB photons by free electrons in the late Universe regenerates large-scale polarisation, giving rise to the so-called \emph{reionisation bump} at $\ell \lesssim 10$ in the $B$-mode power spectrum. The amplitude of this signal at the angular power spectrum level is parametrised by the tensor-to-scalar ratio $r$. 

Detecting the primordial $B$-mode polarisation pattern in the CMB would provide a powerful probe of the inflationary epoch, offering direct evidence of inflationary gravitational waves and valuable insight into the energy scale of the early Universe. Consequently, this detection is a primary goal of several current and forthcoming experiments, including the Simons Observatory (SO) \cite{Ade_2019}, BICEP/Keck \cite{BK2021}, and \textit{LiteBIRD} \cite{LiteBIRD_2023}. The most stringent current upper bound on $r$, obtained from the combination of Planck satellite observations with BICEP2/Keck data, is $r < 0.028$ \cite{2022_Tristram,2023_Galloni}.

Superimposed on the CMB signal are additional Galactic and extragalactic microwave emissions. Within our Galaxy, the dominant sources of polarised emission are thermal dust and synchrotron radiation. Thermal dust emission arises from interstellar dust grains aligned with the Galactic magnetic field, which radiate thermally at high frequencies (above roughly 100 GHz). Synchrotron emission, by contrast, is produced by relativistic cosmic-ray electrons spiralling along magnetic field lines, dominates at low frequencies (below approximately 70 GHz), and exhibits a high degree of polarisation. These polarised Galactic components exceed the expected primordial $B$-mode signal across a broad range of angular scales and therefore pose a significant challenge to its detection \cite{Planck_2014, Planck2015:X, Krachmalnicoff_2016}. Consequently, robust component-separation techniques are essential to extract the weak CMB signal from strong foreground emissions and minimise bias in the estimated value of the tensor-to-scalar ratio, $r$.

The nearly perfect blackbody spectrum of the CMB, described by Planck’s law \cite{1992COBE}, provides a key means of distinguishing it from other foregrounds. In practice, these components can be effectively separated by observing the sky at multiple frequencies, a strategy that has motivated the design of modern CMB experiments such as \textit{COBE} \cite{1992COBE}, \textit{WMAP} \cite{Bennett_2003}, and \textit{Planck} \cite{Planck_2014}.

Several ongoing and planned CMB experiments aiming at the detection of a primordial background of gravitational waves are ground-based and therefore rely on accurate reconstructions of the CMB polarisation field over only a fraction of the celestial sky. Within this experimental framework, the CMB $B$-mode power spectrum can be recovered either by directly fitting the measured multi-frequency power spectra with models that account for foreground emission~\cite{BK2021,Azzoni_2021,2022Vacher}, or by first applying a map-based component-separation pipeline. In view of forthcoming high-sensitivity data, it is essential to develop both approaches in parallel in order to cross-validate the resulting measurements~\cite{Wolz_2024}.

Among the map-based methods, the Internal Linear Combination (ILC) technique~\cite{Tegmark_1996,Eriksen_2004}, implemented in the needlet domain~\cite{Marinucci_2007} and commonly referred to as the Needlet ILC (NILC)~\cite{Delabrouille_2008}, constitutes one of the primary pipelines considered (for instance, Pipeline~B in Ref.~\cite{Wolz_2024}).
Operating in the needlet domain enables the extraction of the cosmological signal from multi-frequency sky observations in a localised and scale-dependent manner. However, residual foreground contamination caused by the complex spatial variability of the spectral energy distributions (SEDs) of Galactic foregrounds can still introduce significant bias in the recovered $B$-mode power spectrum. In this work, we present extensions to the NILC approach aimed at mitigating foreground-induced biases in the estimation of the tensor-to-scalar ratio $r$, and assess their performance in the context of ongoing ground-based CMB experiments. Our methodology introduces two key improvements:
\begin{enumerate}
    \item The incorporation of spectral constraints on foreground emission within the NILC component-separation framework, through a tailored implementation of the constrained moment ILC (cMILC) technique~\cite{Chluba_2017,Remazeilles_2020};
    \item The inclusion of an explicit spectral model for foreground residuals directly in the cosmological likelihood, following the approach proposed in Ref.~\cite{Carones_marg}.
\end{enumerate}

We validate this framework using realistic simulations of SO~\cite{Ade_2019}, and demonstrate that these extensions significantly reduce the bias on the tensor-to-scalar ratio $r$ induced by foreground residuals, while preserving competitive statistical uncertainties. The choice of SO as the instrumental benchmark is motivated by the following reasons: (i) it represents a key and state-of-the-art experimental platform for future ground-based CMB polarisation observations; and (ii) it enables a direct comparison of our results with the baseline forecasts recently produced by the SO collaboration~\cite{Wolz_2024}.

SO is a ground-based experiment located in the Atacama Desert of Chile, which began observations in early 2024. It is designed to measure both CMB temperature and polarisation anisotropies across a wide range of angular scales. The observatory in its nominal configuration comprises one 6-meter Large Aperture Telescope (LAT) and three 0.5-meter Small Aperture Telescopes (SATs). The LAT provides arcminute-scale resolution and will survey roughly $40\%$ of the sky, while the SATs, with an angular resolution of about $0.5^{\circ}$ at $93$GHz, are optimised for large-scale polarisation measurements over an effective sky fraction of about $10\%$. The primary goal of SATs is to constrain the primordial tensor-to-scalar ratio $r$ with a target uncertainty of $\sigma(r) = 0.003$, focusing on the \textit{recombination bump}. Both telescope systems will observe in six frequency bands centred at 27, 39, 93, 145, 225, and 280 GHz, allowing precise separation of Galactic foregrounds from the CMB signal \cite{Ade_2019}.

The structure of this paper is as follows. Section~\ref{sec: compsep} describes the various component separation methods used in this study. In Section~\ref{sec: inputs}, we introduce the simulated data sets employed in our analysis, including Galactic foregrounds, instrumental noise, and CMB signal. Section~\ref{sec: nilc cmilc results} presents our results without the inclusion of foreground spectral templates, as well as the procedure used to infer cosmological parameters from the cleaned CMB maps. Section~\ref{sec: margnalization} details the construction of a spectral model for the foreground residuals, its inclusion in the cosmological likelihood, and the results obtained with this marginalisation procedure. Finally, Section~\ref{sec: conclusion} summarises our findings and discusses their broader implications.

\section{Component Separation} \label{sec: compsep}
At microwave frequencies, sky observations consist of a combination of the CMB signal and various astrophysical foreground emissions, most notably, Galactic synchrotron and thermal dust radiation in polarisation. Synchrotron emission arises from relativistic electrons spiralling along Galactic magnetic field lines, producing a power-law spectrum whose intensity $I_{\nu}$, in the direction of observation $\hat n$, scales with frequency $\nu$ as
\begin{equation}
I_\nu^{\mathrm{sync}}(\hat n) \propto \nu^{\beta_s(\hat n)},
\end{equation}
where $\beta_s$ denotes the synchrotron spectral index \citep{Rybicki:847173}. 

Thermal dust emission, by contrast, originates from interstellar dust grains heated by starlight and is typically modelled as a modified blackbody (MBB) spectrum:
\begin{equation}
I_\nu^{\mathrm{dust}}(\hat n) \propto \nu^{\beta_d(\hat n)} B_\nu(T_d(\hat n)),
\end{equation}
where $\beta_d$ and $T_d$ represent the dust spectral index and temperature, respectively \citep{Finkbeiner_1999, Planck_2018_XI}.


Component separation in CMB data analysis aims to disentangle the CMB signal from various astrophysical foreground emissions. In $B$-mode studies, this task is particularly challenging because the CMB signal is several orders of magnitude weaker than the Galactic emission \cite{Krachmalnicoff_2016}, and the spectral energy distributions (SEDs) of the foregrounds vary significantly across the sky. Several techniques have been developed to address this problem, which can be broadly categorised into \textit{parametric} and \textit{blind} (or \textit{non-parametric}) approaches.

\textit{Parametric methods} fit physically or empirically motivated SED models for each component directly to the multi-frequency data. These techniques jointly estimate the component amplitudes and their spectral parameters, with notable examples including \textsc{Commander}~\cite{Eriksen_2006}, \textsc{FGBuster}~\cite{Stompor_2009}, which operate in pixel space, and the so-called ``cross-spectra" pipelines \cite{Azzoni_2021,2022Vacher} that work at the power spectrum level.
In contrast, \textit{non-parametric approaches} such as template fitting~\cite{Bennett_2003, Planck_2014} and ILC methods~\cite{Tegmark_1996, Eriksen_2004, Vio_2008, Basak_2011} do not rely on explicit foreground models.

In this work, we focus on ILC-based techniques, which have been successfully applied to \textit{WMAP}~\cite{Bennett_2003} and \textit{Planck}~\cite{Planck_2014} data. To extract the CMB signal, the ILC exploits the well-known blackbody spectrum of the CMB and suppresses foregrounds, noise, and systematics by minimising the total variance of the multi-frequency data.
This makes the ILC approach simple, computationally efficient, and robust against modelling uncertainties, motivating its widespread use in CMB data analysis. 

Specifically, we employ three component-separation methods. The first two, blind NILC and semi-blind cMILC, are used to reconstruct CMB maps from the multi-frequency observations, while the third, the Generalised Needlet ILC (GNILC), is applied to obtain cleaned estimates of the foreground emission at each frequency channel. The resulting sets of GNILC maps are then used to construct a spectral template of the foreground residuals remaining after component separation, as detailed later in Section~\ref{sec: marg l(r)}. The principles and implementation details of these methodologies are described in the following subsections.


\subsection{Needlet Internal Linear Combination (NILC)} \label{Nilc section}

The ILC method estimates the CMB signal by linearly combining multi-frequency maps under the assumption that the CMB follows a blackbody spectrum, while the foreground and noise contributions exhibit distinct frequency dependencies. For an input set of multi-frequency maps $X_\nu(\hat{n})$ along the line of sight $\hat{n}$, the ILC estimate of the CMB is given by
\begin{equation}
X^{\mathrm{ILC}}(\hat{n}) = \sum_{\nu} w_\nu(\hat{n}) \cdot X_\nu(\hat{n}),
\qquad \sum_\nu w_\nu(\hat{n})A^{\nu}_{\rm CMB} = 1,
\end{equation}
where the weights $w_\nu(\hat{n})$ are obtained by minimising the variance of the reconstructed map, subject to the constraint that the CMB signal characterised by its SED $A_{\mathrm{CMB}}$ is preserved. This ensures the minimisation of the total contribution from foregrounds and instrumental noise to the overall variance of the solution. When applied in pixel space, the ILC does not account for the fact that instrumental noise dominates on small angular scales (high multipoles), while foreground emission and $1/f$ correlated noise are more prominent on large scales. The Needlet ILC (NILC) addresses this limitation by operating in needlet space \cite{Narcowich_2006, Marinucci_2007, Delabrouille_2008}, a spherical wavelet basis that provides localisation in both the multipole and pixel domains. A set of harmonic filters known as the needlet windows, $b_j(\ell)$ is defined over multipole ranges $\ell$ and satisfies the completeness relation
\begin{equation} \label{eq:b}
\sum_{j=1}^{\infty}\left[b_{j}(\ell)\right]^2 = 1\quad \forall \ell.
\end{equation}
The multi-frequency maps are then filtered in harmonic space using the needlet windows $b_j(\ell)$, producing spherical maps of the so-called needlet coefficients. These coefficients are subsequently combined with minimum-variance weights, optimised independently at each needlet scale $j$. The weights are derived from the local covariance of the needlet coefficients as
\begin{equation} \label{weights} \mathbf{w}_{j}(\hat{n}) = \left[\mathbf{A}_{\rm CMB}^T \mathbf{C}_{j}^{-1}(\hat{n})\,\mathbf{A}_{\rm CMB}\right]^{-1}\mathbf{C}_{j}^{-1}(\hat{n})\,\mathbf{A}_{\rm CMB} \,, \end{equation}
where $\mathbf{C}_{j}(\hat{n})$ denotes the covariance matrix of the needlet coefficients at scale $j$ in the direction $\hat{n}$ and $\mathbf{A}_{\rm CMB}$ is the CMB SED vector. All vectors and matrices are defined in frequency space, with elements corresponding to the different observed frequency channels.
In practice, $\mathbf{C}_{j}(\hat{n})$ is estimated locally by averaging over a circular region surrounding each direction $\hat{n}$. This approach allows the weights $\mathbf{w}_{j}(\hat{n})$ to adapt to local spatial variations in foregrounds and noise across the sky. The size of the averaging domain is chosen to depend on the needlet scale, with smaller $j$ (corresponding to larger angular scales) associated with larger sky regions.

The cleaned CMB estimates obtained at the various needlet scales (i.e.\ across different ranges of angular scales) are then recombined through an inverse needlet transform~\cite{Marinucci_2007}, yielding the final CMB map in real space.

\subsection{Constrained Moment ILC (cMILC)} \label{sec:cMILC}


Any observation of diffuse sky emission with a telescope is inevitably affected by averaging effects along the line of sight and across the instrumental beam. Even if each emitting element follows a well-defined intrinsic SED, these averaging effects distort the observed frequency scaling of the foreground components. Consequently, the effective SED along a given line of sight can be expressed in terms of the moments of the underlying spectral-parameter distributions~\cite{Chluba_2017}.

Within this framework, the foreground intensity $I_\nu(\hat{n})$ in direction $\hat{n}$ can be modelled through a Taylor expansion around a set of pivot spectral parameters (e.g., the synchrotron spectral index $\beta_s$, or the dust temperature $T_d$ and spectral index $\beta_d$):
\begin{equation}
I_\nu(\hat n) = \bar{I}(\hat n)\left[f\!\left(\nu,\bar\beta\right)
+ \sum_i \bigl(\beta_i(\hat n)-\bar\beta_i\bigr)\,\frac{\partial f}{\partial \beta_i}\left(\nu,\bar\beta\right) + \ldots\right],
\label{eq:moment_expansion}
\end{equation}
where $\bar{I}(\hat{n})$ denotes the foreground amplitude, $f(\nu, \bar{\boldsymbol{\beta}})$ is the mean SED evaluated at the pivot parameters $\bar{\boldsymbol{\beta}}$, and the index $i$ runs over the relevant spectral parameters (e.g., $\beta_s$, $T_d$, $\beta_d$). The pivot parameters $\bar\beta_i$ represent the fiducial or mean values of the spectral parameters, typically estimated from global fits or sky averages, around which the SED is linearised. Higher-order terms encode distortions, including those induced by variations of the spectral parameters around their mean values. To mitigate biases from Galactic foreground spectral distortions due to averaging, constrained moment ILC (cMILC) extends the standard NILC by applying, within the weights estimation, additional nulling constraints on independent terms of the foreground moment expansion  ~\cite{Remazeilles_2020}. Therefore, cMILC can better account for spatial variations of the foreground spectral parameters and averaging effects along the line of sight that affect the observed SED.

The cMILC weight vector $\mathbf{w}$ is computed by minimising the variance under multiple constraints:
\begin{equation}
\min_{\mathbf{w}} \; \mathbf{w}^T \mathbf{C} \, \mathbf{w}
\quad\text{subject to}\quad
\mathbf{w}^T \mathbf{A}_{\rm CMB}=1,
\quad \mathbf{w}^T \mathbf{A}_\alpha = 0 \quad (\forall \alpha),    
\end{equation}
where $\mathbf{C}$ is the data covariance matrix and $\mathbf{A}_{\alpha}$ are the homogeneous SED vectors corresponding to selected foreground moments ~\cite{Remazeilles_2011}. Explicitly:
\begin{equation}
\mathbf{w} = \mathbf{e}^T \bigl(\mathbf{A}^T \mathbf{C}^{-1} \mathbf{A}\bigr)^{-1} \mathbf{A}^T \mathbf{C}^{-1},    
\end{equation}
where $\mathbf{A} = [\,\mathbf{A}_\text{CMB},\,\mathbf{A}_1,\,\dots,\mathbf{A}_m\,]   
$ and $\mathbf{e} = [1,0,\dots,0]^T$.

By nulling these moment components, cMILC significantly reduces foreground residual bias compared to NILC. However, each additional nulling constraint limits the effective degrees of freedom available for variance minimisation, resulting in increased reconstruction noise \cite{Remazeilles_2020}. In summary, cMILC achieves a controlled trade‑off such that it deprojects leading foreground spectral-moment modes to suppress bias, while relying on blind variance minimisation for unmodelled or higher-order foreground contributions. 

In principle, different deprojection schemes can be applied at distinct needlet scales, allowing one to balance foreground removal against the corresponding increase in reconstruction noise locally in multipole space. Although not explored in this work, we note that moment deprojection within the ILC framework can be automated through a data-driven approach, which determines the optimal number and subset of moments to deproject, as well as the appropriate degree of deprojection \cite{Carones_ocMILC}.

\subsection{Generalised NILC (GNILC)}
\label{sec: GNILC}
To obtain a reliable estimate of the spatial distribution of the foreground residuals still contaminating the recovered CMB solution, we first require cleaned templates of the input Galactic emission. These are obtained by applying GNILC to the simulated data sets. GNILC exploits both spatial and spectral information to estimate, at each position on the sky, the local dimensionality of the foreground subspace~\cite{Remazeilles_2011}.
By incorporating prior knowledge of the expected CMB and instrumental noise (nuisance) contributions to the total covariance, GNILC can effectively distinguish between modes dominated by foregrounds and those dominated by nuisance components, thereby performing a blind component separation. Specifically, GNILC operates by estimating the local covariance matrix of the observed data, $\widehat{\mathbf{R}}(\hat{n})$, at each sky position $\hat{n}$. This matrix contains contributions from Galactic foregrounds, the CMB, and instrumental noise:
\begin{equation}
    \widehat{\mathbf{R}}(\hat{n}) = \mathbf{R}_\mathrm{fg}(\hat{n}) + \mathbf{R}_\mathrm{cmb}(\hat{n}) + \mathbf{R}_\mathrm{noise}(\hat{n}),
\end{equation}
where we assume negligible cross-correlation among the different components, as expected from the statistical independence of Galactic foregrounds, CMB, and instrumental noise. A prior estimate of the CMB and noise covariance, $\mathbf{R}_{\mathrm{cmb}+\mathrm{noise}}(\hat{n})$, is provided from simulations or theoretical models. The key idea is to whiten the data covariance with respect to this prior:

\begin{equation}
    \mathbf{M}(\hat{n}) = \mathbf{R}_{\mathrm{cmb}+\mathrm{noise}}^{-1/2}(\hat{n}) \, \widehat{\mathbf{R}}(\hat{n}) \, \mathbf{R}_{\mathrm{cmb}+\mathrm{noise}}^{-1/2}(\hat{n}).
\end{equation}
In this whitened space, any eigenvalue of $\mathbf{M}(\hat{n})$ significantly greater than unity indicates a foreground-dominated mode.

The number $m$ of independent foreground modes is determined by minimising the Akaike Information Criterion (AIC):
\begin{equation}
    \mathrm{AIC}(m) = n\cdot\left(2m +  \sum_{i=m+1}^{N_\nu} \left[ \lambda_i - \log \lambda_i - 1 \right]\right),
\end{equation}
where $\lambda_i$ are the eigenvalues of $\mathbf{M}(\hat{n})$, $n$ is the number of data samples used to estimate the covariance, and $N_\nu$ is the number of frequency channels.

Once the foreground subspace dimension is identified, GNILC projects data onto this subspace and performs an internal linear combination to reconstruct the foreground emission. It produces foreground templates at each frequency channel, retaining only modes with variance greater than the predicted contribution from CMB and noise ~\cite{Remazeilles_2011, Olivari_2015}. While some residual CMB and noise contamination may remain, the implemented nuisance subspace deprojection minimises this leakage, allowing GNILC to effectively isolate significant foreground structures while reducing biases from cosmological and instrumental backgrounds.

Cleaned template maps of foreground emission at the different observed frequency channels are constructed using the GNILC method. The central idea is to construct, from simulated data, clean templates of Galactic emission at each frequency channel and to combine them with the component-separation weights obtained from NILC and cMILC. This allows to derive an estimate of the residual foreground contamination at the power spectrum level, which is subsequently used for bias mitigation in the parameter inference described in Section~\ref{sec: margnalization}.

\section{Simulated Data} \label{sec: inputs}

The goal of this work is to demonstrate the ability of the proposed extensions to the standard NILC framework to reduce potential biases in the estimation of the tensor-to-scalar ratio arising from residual foreground contamination in the NILC-reconstructed CMB polarisation signal. This analysis is carried out in the context of ground-based experiments characterised by limited sky coverage and a restricted number of frequency channels. For this purpose, as anticipated in Section~\ref{sec:intro}, we adopt the forecasted instrumental configuration of the SO-SATs~\cite{Ade_2019}.

We apply the NILC and cMILC component-separation pipelines to 300 simulated SO-SAT-like data sets, including contributions from the CMB, Galactic polarised foregrounds, and instrumental noise. Input CMB anisotropies are generated as Gaussian realisations with the harmonic variance given by the power spectrum $C_{\ell}=\left\langle\left|a^{CMB}_{\ell m}\right|^2\right\rangle$. The fiducial CMB power spectrum is computed with \texttt{CAMB}\footnote{\url{https://camb.readthedocs.io/en/latest/}} \cite{CAMB} assuming the best-fit $\Lambda$CDM cosmological parameters as inferred by Planck's 2018 analysis \cite{Planck_cosmopars} and an input tensor-to-scalar ratio $r = 0$ (no primordial gravitational waves) unless stated otherwise. Input CMB maps are generated for each frequency channel, convolved with a Gaussian beam of full width at half maximum ($FWHM$) as listed in Table \ref{tab:noise model}. 
\begin{table}[h]
    \centering
    \begin{tabular}{c|c|c|c|cc}
       \hline
        Frequency  & $FWHM$ & Baseline & Goal  & \multicolumn{2}{c}{Optimistic }\\
        (GHz) & arcmin &$\sigma_{w}$ ($\mu K$-arcmin) &  $\sigma_{w}$ ($\mu K$-arcmin) & $\ell_{knee}$&$\alpha_{knee}$ \\
        \hline
        27 & 91.0 & 46 & 33 & 15 & -2.4\\
        39 & 63.0 &28 & 22 & 15 & -2.4\\
        93 & 30.0 &3.5 & 2.5 & 25&-2.5\\
        145 & 17.0 &4.4 & 2.8 & 25& -3.0\\
        225 & 11.0 &8.1 & 5.5 & 35& -3.0\\
        280 & 9.0 &21 & 14 & 40 & -3.0\\
        \hline
    \end{tabular}
    \caption{The simulation setup for the SO frequency channels incorporates noise characterized by two types: homogeneous map-depth levels ($\sigma_{w}$) and $1/f$ noise, which is defined by a knee multipole ($\ell_{\rm knee}$) and a spectral index ($\alpha_{knee}$). The $FWHM$ indicates the beam size for each frequency.}
    \label{tab:noise model}
\end{table}


To simulate Galactic foreground emission from thermal dust and synchrotron radiation, we employed the \texttt{PySM} (Python Sky Model) package\footnote{\url{https://github.com/galsci/pysm}} \citep{Thorne_2017}. We considered two foreground models. The first, labeled \texttt{d1s1}, combines the \texttt{d1} dust and \texttt{s1} synchrotron prescriptions implemented in \texttt{PySM}. In the \texttt{d1} model, both the dust spectral index $\beta_{d}$ and dust temperature $T_d$ vary across the sky, with spatial templates derived from the \textit{Planck} data using the \textsc{Commander} code \citep{Planck2015:X}. The synchrotron emission, described by the \texttt{s1} model, follows a power-law spectrum with a spatially varying but frequency-independent spectral index $\beta_s$. The synchrotron spectral-index map was obtained by combining the Haslam 408\,MHz data with the 7-year \textit{WMAP} 23\,GHz data \citep{MivilleDeschenes_2008}.
The second model, \texttt{d10s5}, is more complex as it features greater spatial variability in the spectral parameters of both dust and synchrotron radiation. It also includes the injection of non-Gaussian small-scale features in the templates and spectral parameters by means of the \texttt{logpoltens} formalism \cite{PySM2025}. Polarised extragalactic sources are excluded because they are anticipated to be significant at angular scales smaller than those primarily relevant for inflationary B-mode science ($\ell > 100$) ~\cite{Puglisi_2018}.

\begin{figure}
    \centering
    \includegraphics[width=0.6\linewidth]{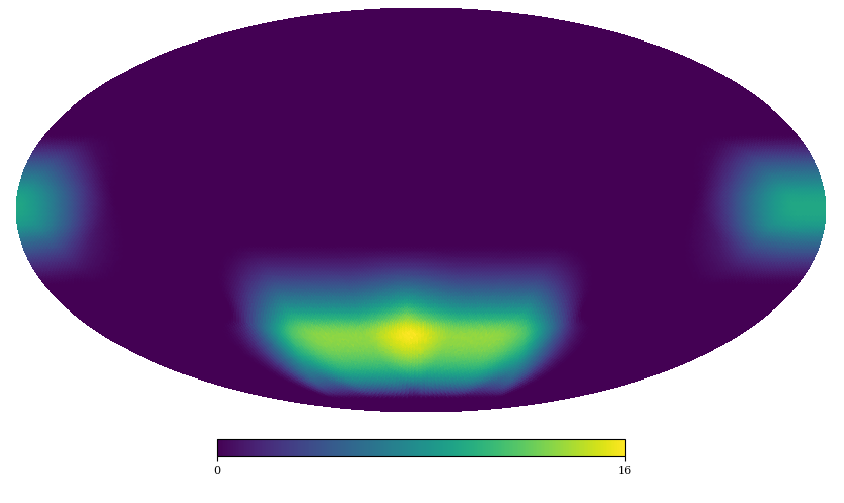}
    \caption{SO-SAT forecasted hit-count map shown in equatorial coordinates.}
    \label{fig: hit-count map}
\end{figure}

Input noise maps for the $Q$ and $U$ Stokes parameters are generated as Gaussian realisations drawn from a model power spectrum that includes both white-noise and $1/\ell$ components:
\begin{equation}
   N_{\ell}=N_{w}\left[1+\left(\frac{\ell}{\ell_{knee}}\right)^{\alpha_{knee}}\right],
\end{equation}
where $N_{w}$ is the white noise spectral amplitude, while $\alpha_{knee}$ and $\ell_{knee}$ characterise the contribution from correlated noise. The white-noise level is related to the map depth $\sigma_{w}$ through
\[
N_{w} = \frac{\pi\, \sigma_{w}}{180 \times 60},
\]
where the factor converts the depth from $\mu K$-arcmin to steradians.
The shape of the $1/\ell$ term is derived from the expected spectral transfer function, which characterises the mode suppression induced by time-ordered data (TOD) filtering. In real observations, this filtering procedure is applied to mitigate atmospheric fluctuations and other large-scale systematics.
Table~\ref{tab:noise model} summarises the specifications adopted to simulate both the white and $1/\ell$ noise components. The baseline noise scenario of our analysis follows the \emph{baseline-optimistic} noise model of~\cite{Wolz_2024}, which assumes nominal white-noise levels and an optimistic characterisation of the $1/\ell$ component. For selected cases, we also consider an improved noise model featuring a lower white-noise floor, corresponding to the \emph{goal-optimistic} configuration described in~\cite{Wolz_2024}. 
Finally, the forecasted hit-counts map is used to weight the noise realisations on a pixel-by-pixel basis, thereby accounting for the non-uniform sky coverage arising from the scanning strategy. The adopted hit-counts map, shown in Figure \ref{fig: hit-count map}, results from the planned scanning strategy for SATs and is consistent with the one presented in \cite{Wolz_2024}. 

To simulate CMB, noise and foreground maps with the specifications and anticipated observational capabilities of the SO \cite{Ade_2019}, we used the publicly available Python package \texttt{BROOM}\footnote{\url{https://github.com/alecarones/broom.git}} (paper in preparation). All the 300 input data-sets are simulated adopting the HEALPix \cite{Gorski_2005} pixelization scheme with a resolution parameter of $N_{\texttt{side}} = 128$ and then brought to a common angular resolution corresponding to a Gaussian beam with $FWHM=30$ arcmin.  Since we are targeting primordial $B$ modes for which the expected signal is confined to $\ell < 100$, this resolution is adequate to capture all the relevant scales. 

\section{Component-separation results} \label{sec: nilc cmilc results}

This section presents the results obtained from the application of NILC and cMILC to simulated SO-SAT-like data. The component-separation pipelines are implemented using the \texttt{BROOM} package. 

In this analysis, for simplicity, the inputs to the component-separation routines are $B$-mode maps obtained from the full-sky harmonic transformation of simulated $Q/U$ maps, followed by the application of an appropriate sky mask prior to component separation. Such full-sky harmonic reconstruction is not feasible for real observations owing to the limited sky coverage of ground-based experiments. We therefore verified that consistent results are obtained when analysing the masked $Q/U$ maps directly. In this verification, the masking strategy applied to the simulated data follows the approach described in~\citep{Wolz_2024}. The mask is constructed from the SO-SAT forecasted hit-count maps by normalising it to its maximum value, and subsequently, for power spectrum computation, apodising it with a 15-degree $C1$ kernel to reduce edge-induced mode coupling and $E-B$ leakage.
In realistic analyses, where residual $E$–$B$ leakage may affect the input data, this effect can be mitigated at the map level prior to component separation using the techniques presented in~\citep{Carones_nilc}, which are implemented in \texttt{BROOM}.

\begin{figure}[htbp]
\centering
\includegraphics[width=.5\textwidth]{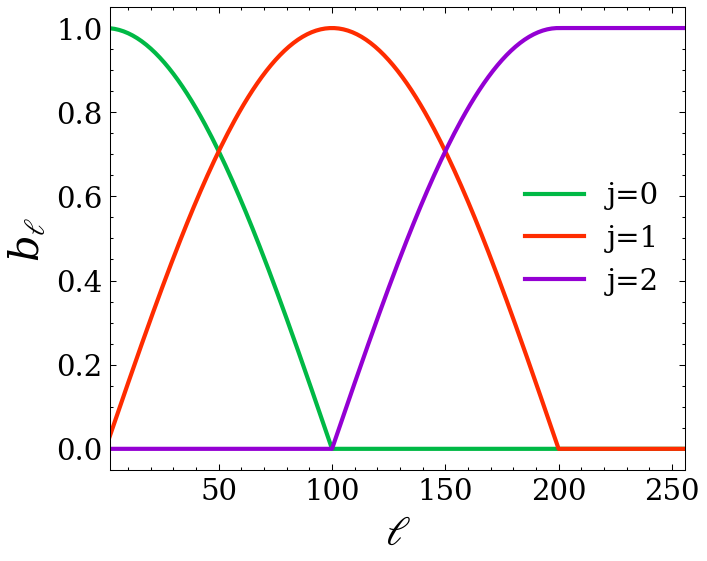}
\caption{Cosine needlet bands used for needlet filtering with peaks at $\ell_\mathrm{peaks} = [0, 100, 200]$.}
\label{fig: cosine needlets}
\end{figure}

To mitigate the effect of uneven sky coverage in component separation, we weight the input needlet maps with the SO hit-count map prior to the covariance computation. This ensures that regions of the sky with denser coverage receive greater statistical weight. For NILC and cMILC analysis, the same needlet configuration (Figure~\ref{fig: cosine needlets}) is used, with cosine harmonic bands centred at multipoles $\ell_\mathrm{peaks} = [0, 100, 200]$. Cosine needlets have been commonly used in CMB data analysis \cite{Planck2018_compsep, Ade_2019}. For the cMILC implementation, we deproject the zeroth moment (i.e., the average spectral behaviour) of both dust and synchrotron components in the first two needlet bands, corresponding to the largest angular scales. For the third needlet band, which probes smaller angular scales, only the zeroth moment of dust emission is deprojected. This configuration strikes a balance between effective foreground mitigation and noise control, as fully deprojecting both dust and synchrotron at high multipoles would otherwise lead to excessive noise amplification. The chosen deprojection scheme was optimised \textit{a posteriori} by examining the residuals from the component-separation outputs and was kept fixed across all the cases analysed in this work. Nonetheless, such an optimisation could have been obtained automatically through the data-driven pipeline presented in~\cite{Carones_ocMILC}.

\subsection{Angular power spectra} \label{sec: angular power spectra}
To assess the performance of the component separation techniques, we estimate the residual contamination from foregrounds and noise in the cleaned CMB B-mode maps. Following the methodology presented in \cite{Carones_nilc}, we compute the cleaned CMB map for each simulated multi-frequency dataset using Equation \eqref{weights}. We then apply the derived weights, denoted as $\mathbf{w_i}$, to the noise-only and foreground-only simulations. This process allows us to obtain estimates of the noise and foreground residuals associated with each cleaned map. For each component separation method, we analyse the average power spectrum of these residuals across all simulations.

The spectra are estimated using the \texttt{MASTER} algorithm, implemented into the \texttt{NaMaster} Python package\footnote{\hyperlink{https://namaster.readthedocs.io/en/latest/}{https://namaster.readthedocs.io/en/latest/}} utilised to account for masking and beam effects \cite{Hivon_2002,  2005Polenta}. A bin size of \(\Delta \ell = 10\) is used for power spectrum extraction \cite{Alonso_2019}. Binning the power spectrum allows for a robust power spectrum derivation in such a small fraction of the sky, making the coupling matrix invertible and its inverse numerically stable. The chosen binning scheme adheres to the one employed in \cite{Wolz_2024}. The mask used for power spectrum computation in this analysis is based on the hits count map released by the SO Collaboration \cite{Ade_2019}, and further apodized with a $C1$ kernel of size 15 degrees \cite{Grain2009}.

\begin{figure}[htbp]
\centering
\includegraphics[width=.48\textwidth]{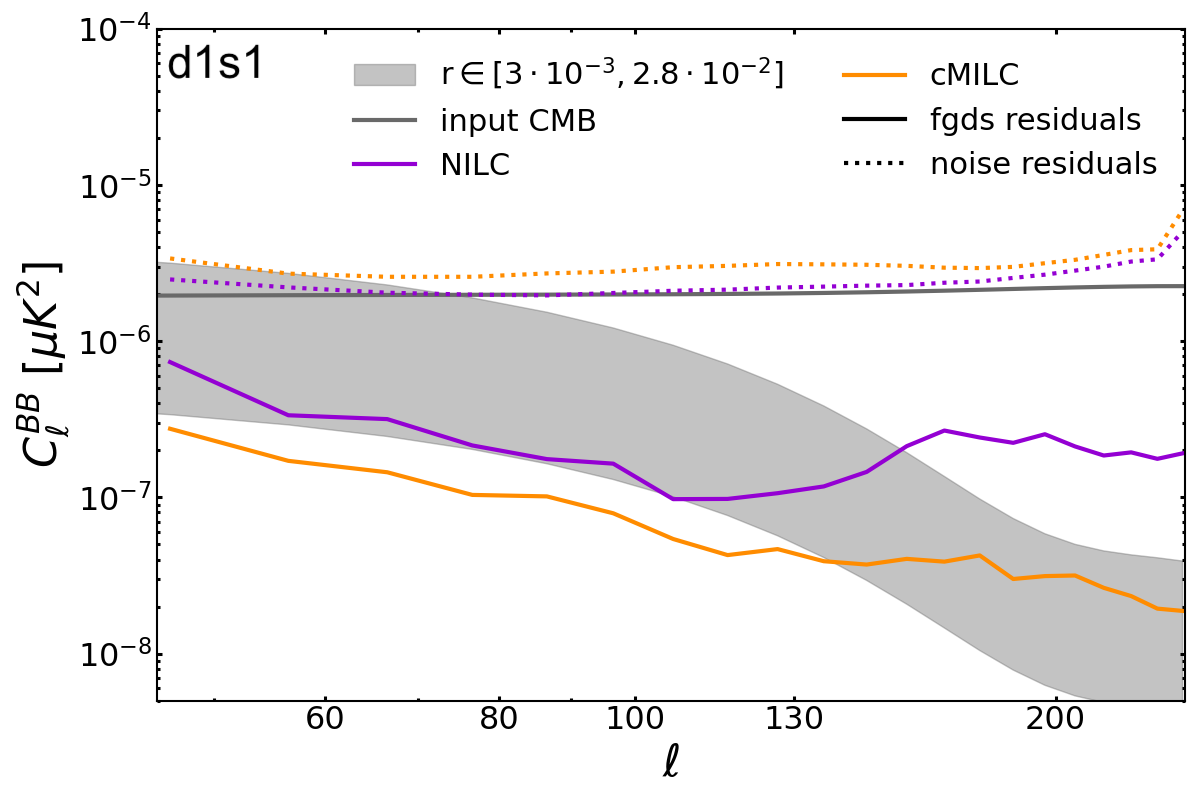}
\includegraphics[width=.48\textwidth]{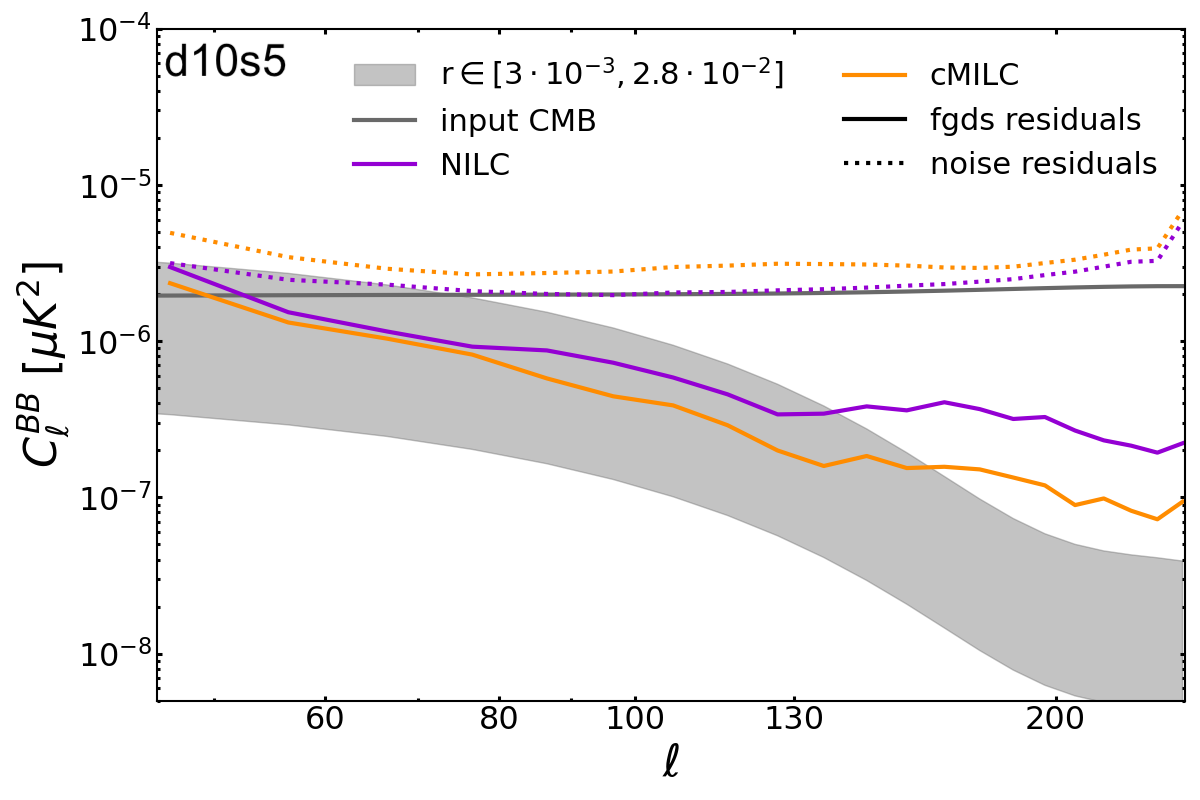}
\caption{$B$-mode angular power spectra of noise (dotted lines) and foreground residuals (solid lines) for NILC (violet) and cMILC (orange). All angular power spectra represent an average over 300 simulations. A uniform binning with $\Delta \ell=10$ is adopted. The average input CMB (with $r=0$) power spectrum obtained after applying the same masking to 300 simulated CMB-only maps, which contain lensing, is shown with the grey solid line. Results are reported for two foreground scenarios: \texttt{d1s1} (left panel) and \texttt{d10s5} (right panel), and assuming the baseline optimistic noise model. The grey shaded area indicates the amplitude range targeted by SO for the primordial tensor signal: $r \in [0.003, 0.028]$.} \label{fig: spectra}
\end{figure}

The average power spectra of the foreground and noise residuals in the reconstructed CMB $B$-mode signal, obtained from the NILC and cMILC component-separation pipelines, are shown in Figure~\ref{fig: spectra}. Results are reported for both the \texttt{d1s1} and \texttt{d10s5} scenarios. As expected, the \texttt{d10s5} case exhibits larger foreground residuals compared to \texttt{d1s1}. This is due to the increased spatial complexity of the dust and synchrotron spectral parameters in \texttt{d10s5}, which makes foreground subtraction more challenging. Conversely, only a marginal increase in reconstruction noise is observed between the scenarios.
In both cases, we find that including the deprojection of foreground statistical moments in the weight estimation leads to a significant reduction of Galactic contamination across all the considered angular scales and foreground models. This improvement is accompanied by a mild increase in the level of noise residuals for cMILC. Overall, for all considered cases, we observe a spectral amplitude of the foreground residuals at a comparable level of values of the tensor-to-scalar ratio targeted by SO-SATs, therefore anticipating residual bias in the inferred cosmological parameters.

\begin{figure}[htbp]
\centering
\hspace{-0.2 cm}\includegraphics[width=.635\textwidth]{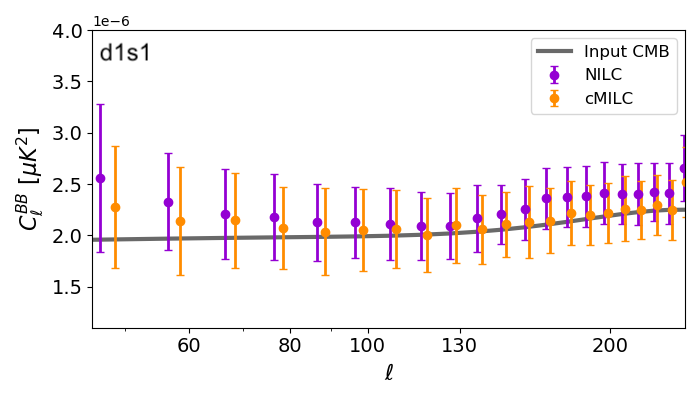 }
\includegraphics[width=.62\textwidth]{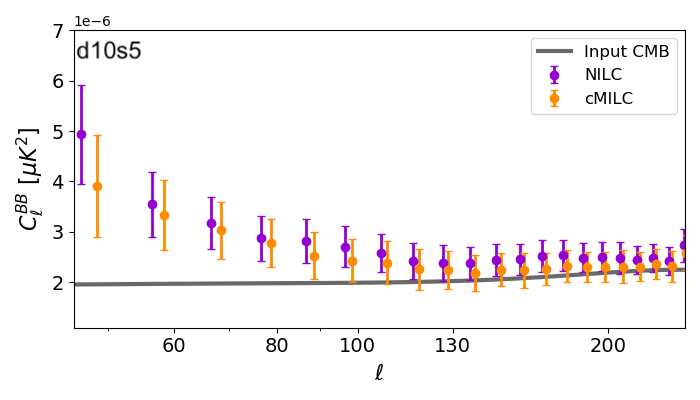}
\caption{Averaged reconstructed CMB $B$-mode power spectra obtained from 300 simulations using the NILC (violet) and cMILC (orange) pipelines for the \texttt{d1s1} (upper panel) and \texttt{d10s5} (lower panel) foreground models. The solid grey line indicates the input CMB spectrum for $r = 0$. A uniform binning with $\Delta \ell=10$ is adopted. Colored markers represent the mean recovered bandpowers, with error bars showing the statistical dispersion across simulations. Data points have been debiased for reconstruction noise by subtracting the mean angular power spectrum of noise residuals obtained in each case.}
 \label{fig: cmb bias}
\end{figure}

Figure~\ref{fig: cmb bias} shows the reconstructed CMB $B$-mode power spectra obtained using the NILC and cMILC pipelines for the \texttt{d1s1} and \texttt{d10s5} foreground models. Data points and error bars correspond to the mean and standard deviation over 300 simulations, respectively. The noise bias has been removed by subtracting from each data point the average angular power spectrum of the noise residuals obtained in the corresponding case. Figure~\ref{fig: cmb bias} confirms the trends already highlighted in Figure~\ref{fig: spectra}: (i) foreground residual bias is more significant for the \texttt{d10s5} scenario and, overall, larger for NILC than for cMILC, with the latter exhibiting only a mild increase in the error bar width due to higher reconstruction noise; and (ii) as expected, the most prominent bias appears at low multipoles. In addition, for NILC, we observe a mild increase in the residual bias towards higher multipoles consistent with previous findings (e.g., Fig. 7 of \cite{Wolz_2024}). This effect can be attributed to beam-induced correlations between neighbouring pixels on small angular scales, which are not explicitly accounted for in the local covariance estimation underlying NILC.
\subsection{Cosmological Inference}\label{sec: Likelihood}

To derive cosmological constraints from the cleaned CMB $B$-mode maps, we adopt a fiducial Gaussian likelihood approximation~\cite{Hamimeche_2008}. Although the CMB power spectrum is not strictly Gaussian distributed on a multipole-by-multipole basis, particularly at large angular scales, the use of binned power spectra (bandpowers) effectively averages over multiple modes. According to the Central Limit Theorem, this averaging renders the distribution of the bandpowers approximately Gaussian~\cite{Bond1997wr}. The adopted log-likelihood is therefore expressed as
\begin{equation}
-2\ln\mathcal{L} \;=\; \sum_{\ell_b,\ell_b'} 
\Delta C_{\ell_b}\; \mathbf{M}^{-1}_{\ell_b\ell_b'}\; \Delta C_{\ell_b'},
\label{eq: gaussian likelihood}
\end{equation}
where $\mathbf{M}^{-1}_{\ell_b \ell_b'}$ is the inverse covariance matrix of the recovered $BB$ power spectrum, estimated from simulated realizations, and $\Delta C_{\ell_b}$ is defined as the difference between the average reconstructed spectrum after component separation, $\widehat{C}_{\ell_b}^{BB,\mathrm{out}}$, and the theoretical model, $C_{\ell_b}^{\mathrm{model}}$:
\begin{equation}
\Delta C_{\ell_b} \;=\; \widehat{C}_{\ell_b}^{BB,\mathrm{out}} - C_{\ell_b}^{\mathrm{model}}(r,A_{ lens}),
\label{eq:deltaC_nomarg}
\end{equation}
Here, $\ell_b$ denotes bandpowers. The model power spectrum is written as
\begin{equation}
C_{\ell_b}^{\mathrm{model}}(r,A_{\rm lens}) \;=\; r\,C_{\ell_b}^{\mathrm{prim}}(r=1) \;+\; A_{\rm lens}\,C_{\ell_b}^{\mathrm{lens}} \;+\; C_{\ell_b}^{\mathrm{nres}},
\label{eq: model}
\end{equation}
where $C_{\ell_b}^{\mathrm{prim}}(r=1)$ is the primordial tensor $BB$ spectrum for $r=1$, 
$C_{\ell_b}^{\mathrm{lens}}$ is the lensing contribution (for $A_{\mathrm{lens}}=1$), 
and $C_{\ell_b}^{\mathrm{nres}}$ represents the residual noise power spectrum after component separation. 
To ensure consistency with the observed bandpowers, both the primordial ($C_\ell^{\mathrm{prim}}$) and lensing ($C_\ell^{\mathrm{lens}}$) fiducial spectra are processed through the same steps as the data, including mode coupling, binning, and multiplication by the inverse of the binned mode-coupling matrix \cite{Alonso_2019}. We note that, in the case the covariance matrix were parameter-dependent, an additional non-constant term including the logarithm of the determinant of the covariance should appear in the likelihood in Equation \ref{eq: gaussian likelihood}. However, in our analysis, the covariance matrix is kept fixed and is estimated assuming the same fiducial model adopted in the simulations. 

The parameters $r$ and $A_{lens}$ are then jointly sampled using a Markov Chain Monte Carlo (MCMC) approach implemented with the \texttt{emcee} sampler. We adopt uniform priors $r \in [-1, 1]$\footnote{Although negative values of $r$ are not physically meaningful, we allow them in the sampling process to ensure unbiased exploration of the parameter space around $r = 0$ and to maintain the approximate Gaussian shape of the posterior in the case of null detections.} and $A_{ lens}>0$. The averaged power spectrum over 300 simulations from the cleaned CMB $B$-mode maps is used to derive the posteriors of $r$ and $A_{\mathrm{lens}}$ using Equations~\ref{eq: gaussian likelihood} and~\ref{eq: model} for both NILC and cMILC. We restrict the fit to the multipole range $32 < \ell \leq 256$, which corresponds to the scales most sensitive to primordial gravitational waves and accessible to the SO-SAT observations \cite{Ade_2019}. Quantitative results are reported in Table~\ref {tab: results}.

The parameters are obtained from the marginal posterior distributions, with the reported values corresponding to the maximum a posteriori estimate of the full posterior, while the quoted uncertainties are given by the 68\% credible intervals of the corresponding marginal posterior distributions.
Both NILC and cMILC yield estimates of $r$ with best-fit values around $1\times10^{-3}$, hence slightly biased by residual foreground contamination given the assumed input value $r_{\mathrm{in}} = 0$ and an uncertainty of $\sigma(r) \simeq 3\times10^{-3}$ for the baseline \texttt{d1s1} model. These results are consistent with the reference constraint on $r$ obtained for an analogous simulated SO-SATs data set in \cite{Wolz_2024} (Table~4), thereby providing a cross-validation of different NILC implementations. The result obtained from the NILC outputs, already reported in Ref.~\cite{Wolz_2024}, provides the main motivation for exploring potential extensions aimed at reducing the impact of residual $B$-mode foreground systematics in ILC solutions derived from the analysis of data collected by ground-based experiments. For the more complex \texttt{d10s5} foreground model, the bias on $r$ increases to the order of $\sim10^{-2}$, with only a 15\% increase in the uncertainty on \textit{r}. In the NILC case, we observe a significant foreground-induced bias not only in $r$ but also in the lensing amplitude $A_{\mathrm{lens}}$. When using cMILC, biases are partially mitigated for both parameters, particularly for $A_{\mathrm{lens}}$, demonstrating the method’s improved capability to handle complex foregrounds. As expected, this improvement comes at the cost of a modest increase in the statistical uncertainty, owing to the higher reconstruction noise introduced by the additional constraints.

Despite this progress, both methods still display residual bias, especially in the more complex foreground regime. This suggests that blind or semi-blind component separation alone is not enough to completely mitigate the impact of foregrounds on CMB $B$-mode cosmological inference. Therefore, in the next section, we present a further step forward in the data analysis, which enables us to properly eliminate the observed foreground-induced biases on cosmological parameters.

\begin{figure}
    \centering
    \includegraphics[width=1\linewidth]{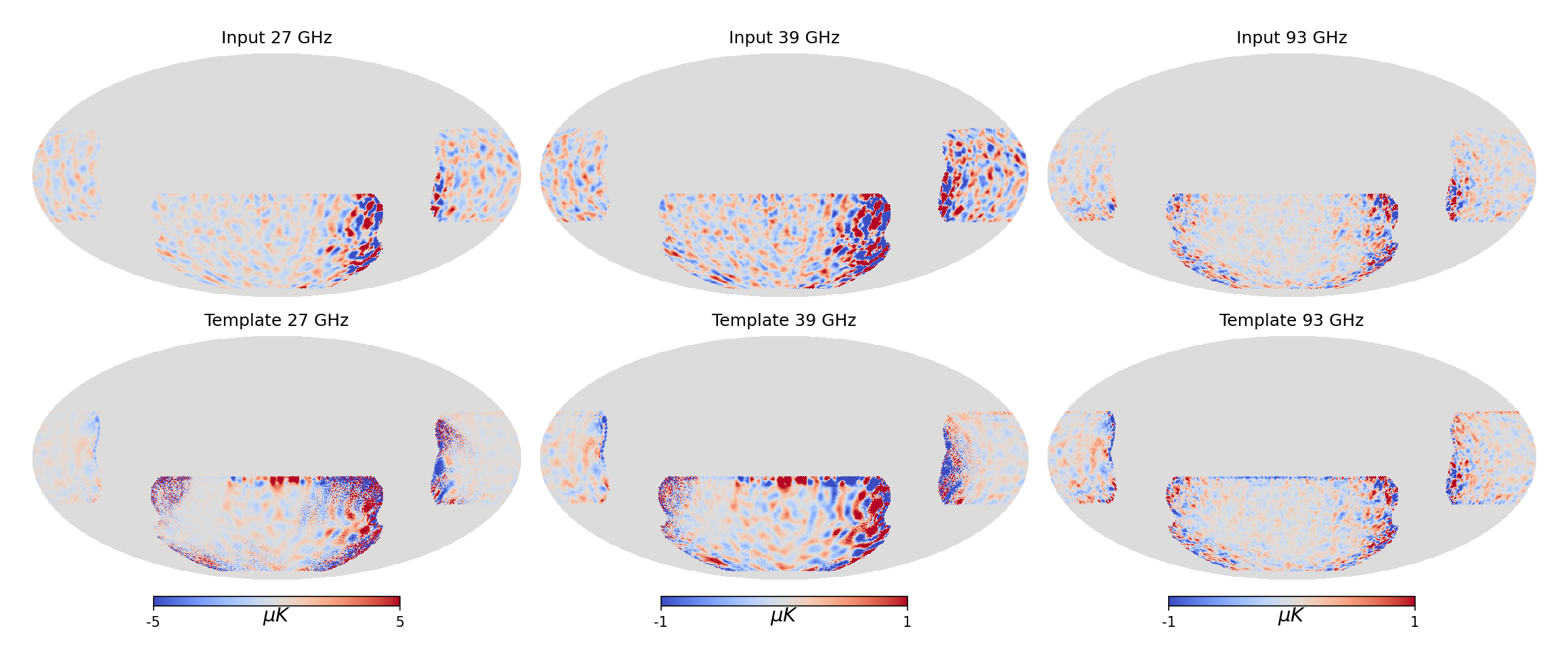} \\
    \qquad
    \includegraphics[width=1\linewidth]{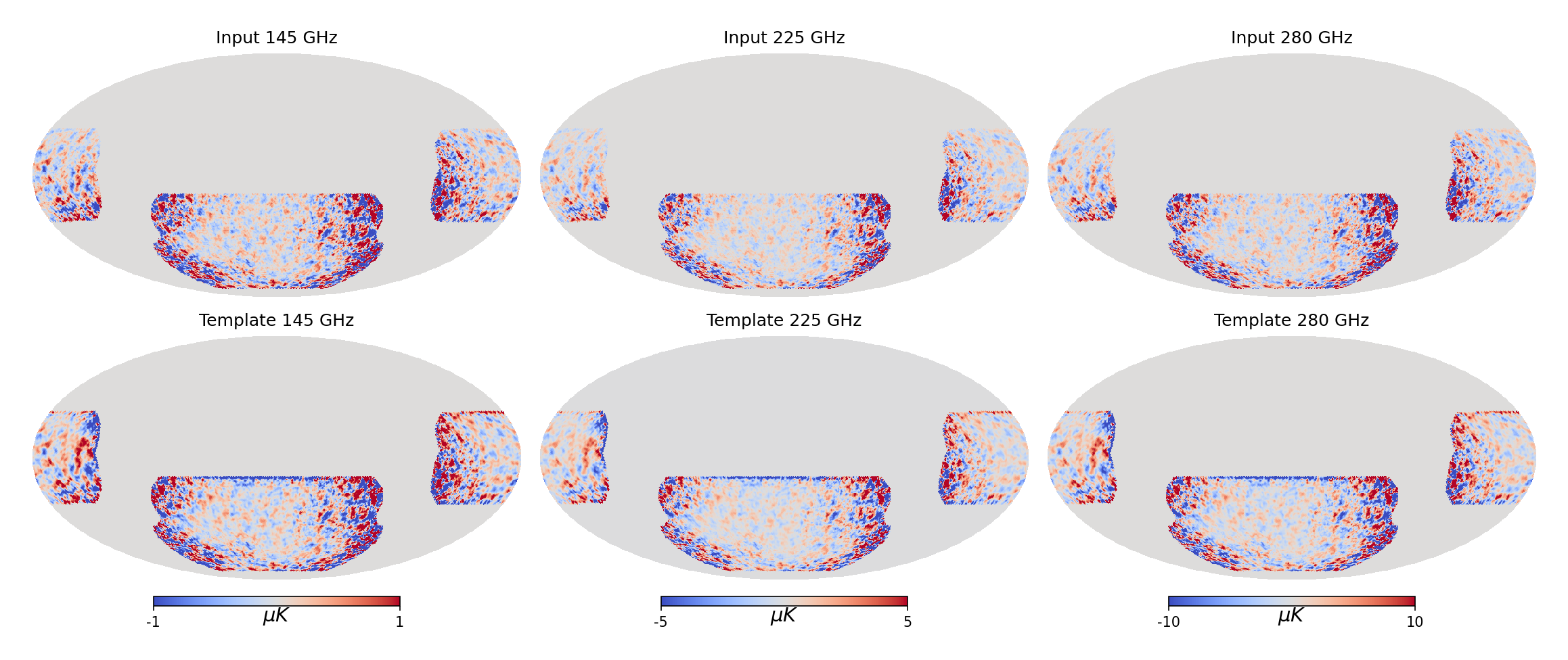}
    \caption{$B$-mode maps of Galactic foregrounds, shown either as simulation inputs (first and third rows) or as reconstructed with GNILC (second and fourth rows), for the \texttt{d1s1} foreground model at the six SO frequency channels. All maps are displayed in units of $\mu$K.
}

    \label{fig: template maps}
\end{figure}
\section{Marginalisation of foreground residuals} \label{sec: margnalization}
In this section, we describe the procedure used to derive a power-spectrum template for the foreground residuals contaminating the observed $B$-mode signal after component separation. This template is subsequently incorporated into the likelihood model (see Equation~\ref{eq: model}) to account for, and mitigate, the bias on the inferred tensor-to-scalar ratio $r$ induced by residual foreground contamination (see Table~\ref{tab: results} for results). The methodology used to reconstruct Galactic emission across the SO frequency bands is described in Section~\ref{sec: GNILC}. 
The procedure to derive a template of component-separation foreground residuals,
and its incorporation into the spectral model are presented in Section~\ref{sec: marg l(r)}, while the corresponding results are discussed in Section~\ref{sec: results with (m)}.

\subsection{Template marginalisation} \label{sec: marg l(r)} 

\begin{figure}[tbp]
\centering
\includegraphics[width=1\textwidth]{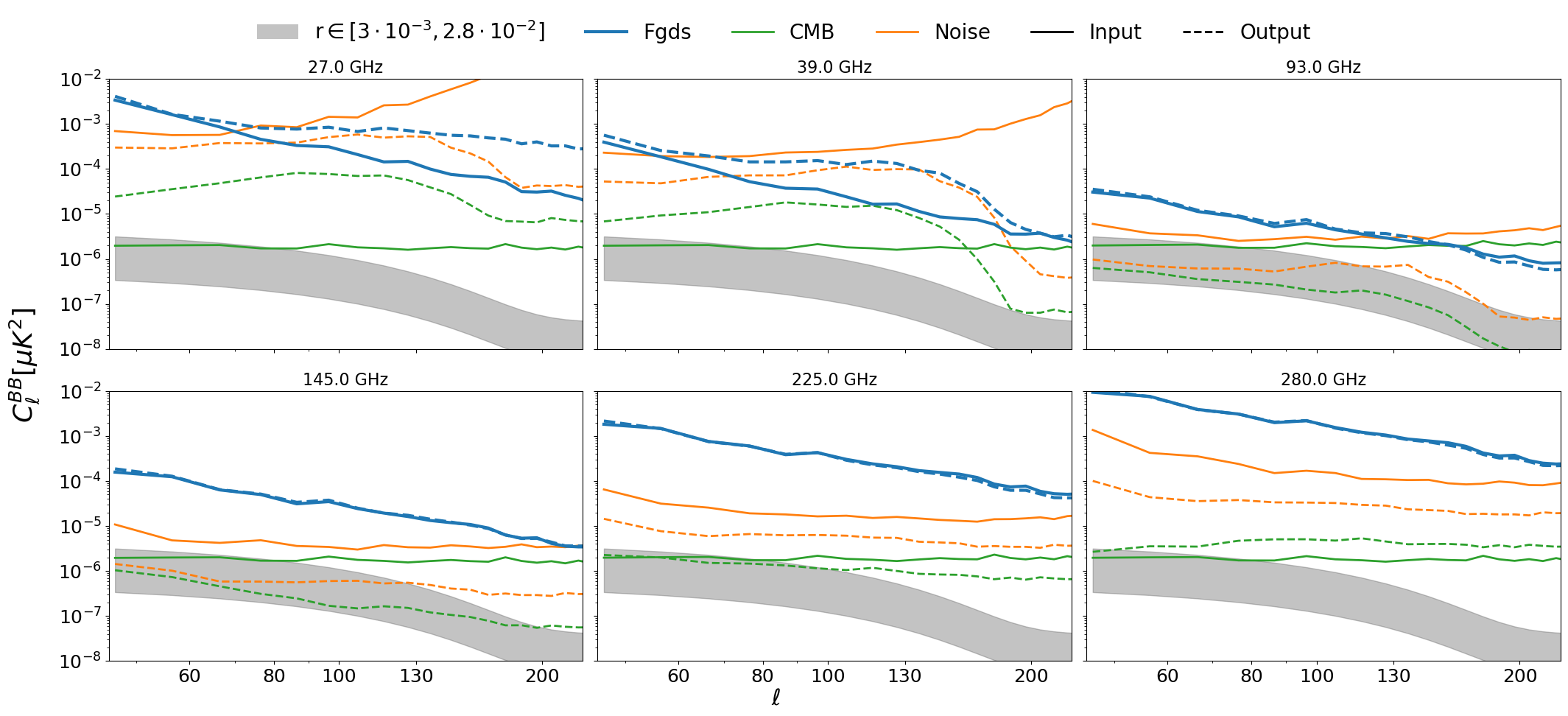}\\
\includegraphics[width=1\textwidth]{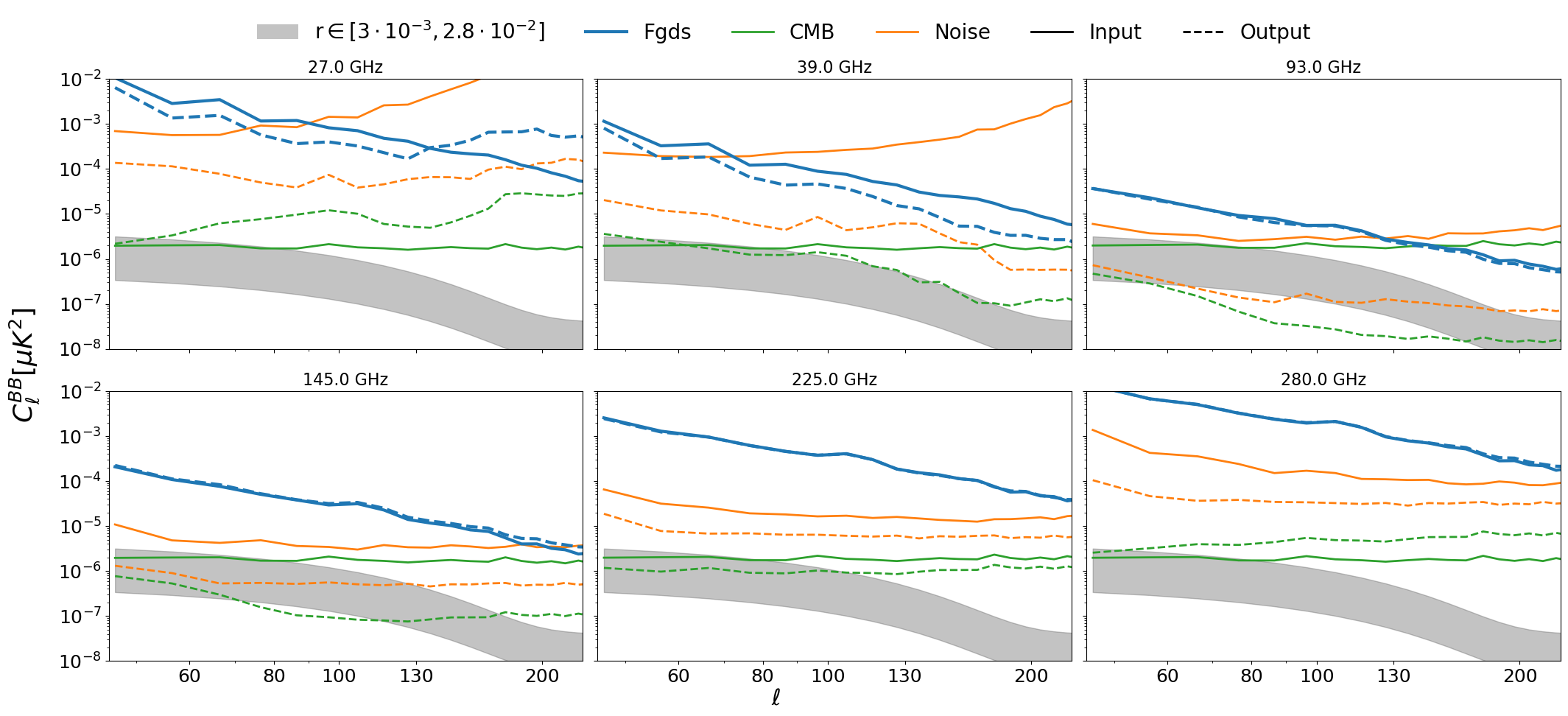}
\caption{Blue solid and dashed lines show the $B$-mode angular power spectra of input and GNILC-reconstructed foregrounds. Corresponding contributions from CMB (green) and instrumental noise (orange) in input data (solid) and GNILC outputs (dashed) are reported. Results are shown for all SO frequency channels and for two foreground scenarios: \texttt{d1s1} (top panel) and \texttt{d10s5} (bottom panel), and assuming the baseline optimistic noise model. The grey shaded region indicates the amplitude range targeted by SO for the primordial tensor signal, corresponding to $r \in [0.003, 0.028]$.} \label{fig: templates}
\end{figure}

The foreground maps derived from the application of GNILC to our multi-frequency simulations are used to construct the template of foreground residuals left by component separation in the CMB solution. Figure~\ref{fig: template maps} shows the GNILC-reconstructed foreground maps (for visualisation purposes, just for the \texttt{d1s1} model) at the six SO frequency channels, alongside the corresponding input simulated foreground maps. 
The templates closely reproduce the spatial features of the input foregrounds, although minor differences remain, particularly at the lowest frequencies (27 and 39 GHz). In addition, the reconstructed maps contain residual CMB and instrumental noise, which is especially evident in the low-frequency channels and in regions near the edges of the observed patch, where instrumental noise becomes more prominent.


Figure~\ref{fig: templates} displays the angular power spectra of the input maps, along with the GNILC-derived foreground templates and the individual components (CMB, noise) that contribute to the foreground templates. These CMB and noise residuals are computed by applying the GNILC-derived weights to the CMB-only and noise-only simulations.  As already suggested by the map-based comparison, the power-spectrum analysis further highlights the difficulty of GNILC in accurately recovering the structure and amplitude of the input foregrounds at low frequencies (dominated by synchrotron emission), while demonstrating its excellent performance in reconstructing the thermal dust component at higher frequencies, with substantially reduced contamination from the CMB and instrumental noise. Specifically, we highlight the appearance of spectral bumps in the CMB and noise residuals of the GNILC low-frequency channels. These features arise from the choice of preserving $m+1$ modes in the GNILC reconstruction within the second and third needlet bands. However, this additional mode is not foreground-dominated at all frequencies. As a consequence, while the adopted strategy improves the reconstruction of dust emission on intermediate angular scales, it does so at the expense of introducing these artefacts in the synchrotron-dominated channels.

The obtained multi-frequency cleaned foreground maps are then combined with the weights of the NILC and cMILC solutions, thus providing as output a map-based template of the residual foreground leakage in the reconstructed CMB signal.
In parallel, simulations of instrumental noise are propagated through the same pipeline to estimate the noise contribution to the residual templates. The final spectral model of the foreground contamination $C_{\ell_b}^{\mathrm{FTres}}$ is then obtained as the difference between the template spectrum and the average spectrum of the noise residuals. Further details of this procedure can be found in~\cite{Carones_marg}.

\begin{figure}
    \centering
    \includegraphics[width=1\linewidth]{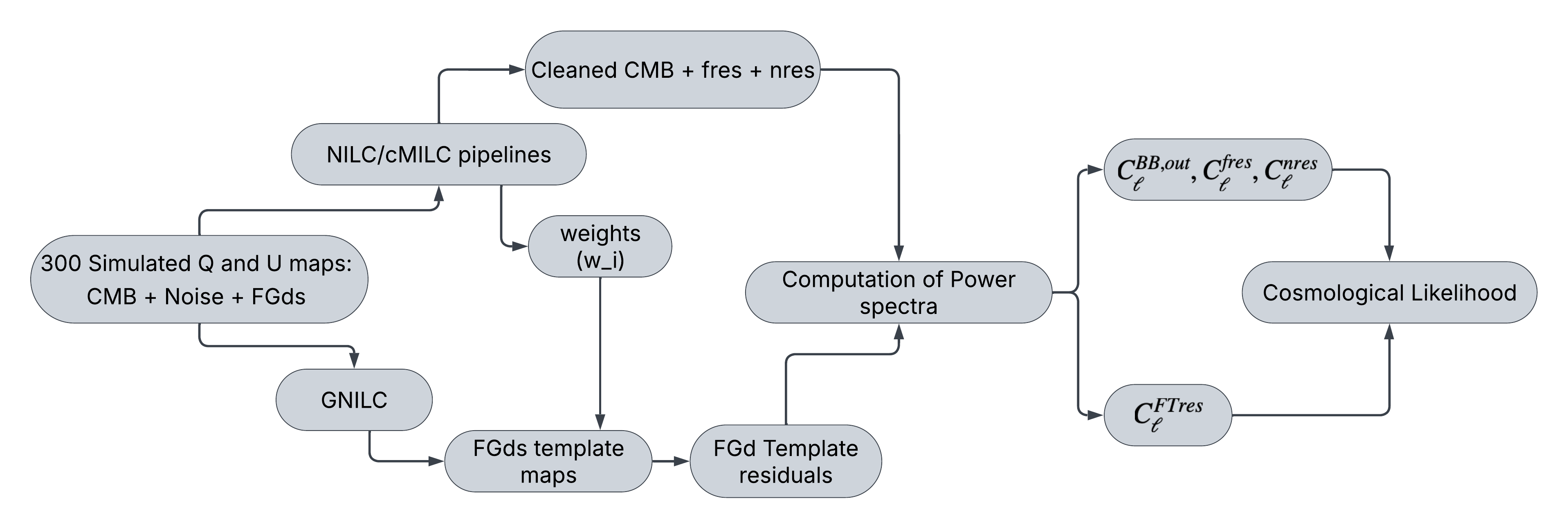}
    \caption{Schematic overview of the procedure used for CMB component separation and cosmological parameter estimation. Starting with 300 simulated Q and U input maps containing CMB, noise, and foregrounds, GNILC is applied to build foreground template maps. These templates are then combined with the NILC/cMILC weights to compute the residual foreground leakage. The power spectra are computed for the component-separation pipeline output maps and subsequently used in the likelihood function in Equation \ref{eq: model temp} to infer cosmological parameters.}
    \label{fig: flowchart}
\end{figure}

To account for the biases observed in results reported in Table~\ref{tab: results} and induced by residual foregrounds contaminating the observed power spectrum, we include an additional term in the model corresponding to the estimated angular power spectrum of the foreground residuals template $C_{\ell_b}^{\mathrm{FTres}}$. In the spectral model, this template is multiplied by an amplitude parameter $a_\text{temp}$, which is sampled together with $r$ and $A_\text{lens}$ to account for any mismatch in the amplitude between template and actual foreground residuals:
\begin{equation}
C_{\ell_b}^{\mathrm{model}}(r,A_{\rm lens},a_{\rm temp}) \;=\; r\,C_{\ell_b}^{\mathrm{prim}}(r=1) 
+ A_{\rm lens}\,C_{\ell_b}^{\mathrm{lens}} + C_{\ell_b}^{\mathrm{nres}} + a_{\rm temp}\,C_{\ell_b}^{\mathrm{FTres}}.
\label{eq: model temp}
\end{equation}
The joint posterior is sampled over \((r, A_{\rm lens}, a_{\rm temp})\) using the same MCMC framework, with uniform priors $r \in  [-1, 1]$, $a_{\rm temp} \in (-50, 50)$ and $A_{\rm lens}>0$, and then the tensor-to-scalar ratio estimate is obtained marginalizing over both lensing and foreground contamination. Therefore, throughout the remaining part of the manuscript, we refer to this case as \textit{marginalisation} case.
A schematic flowchart summarising the complete procedure proposed and implemented in this work, as applied to each simulation, is shown in Figure \ref{fig: flowchart}. We note that similar attempts in deprojecting foreground-residual templates for parametric component-separation pipelines have already been studied and introduced in the literature \cite{2019-Josquin,LiteBIRD_2023, Wolz_2024}.

\begin{figure}[tbp]
\centering
\includegraphics[width=.49\textwidth]{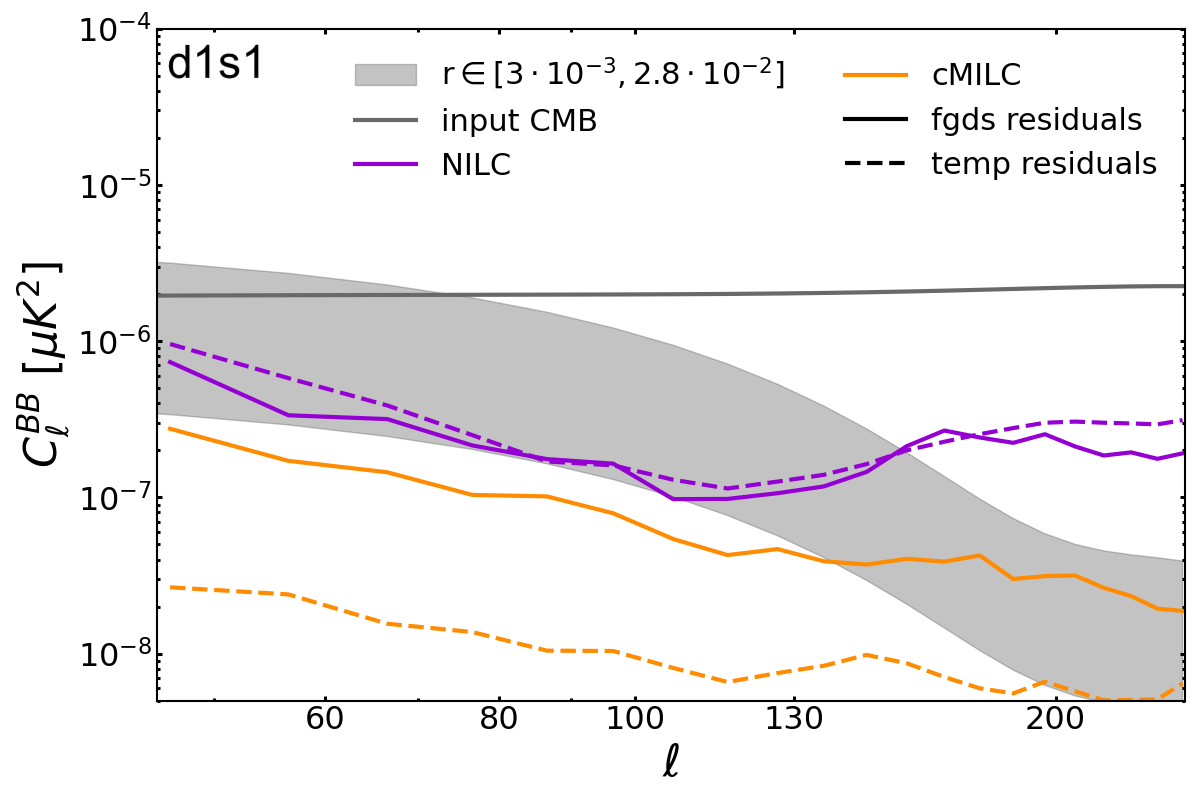}
\includegraphics[width=.49\textwidth]{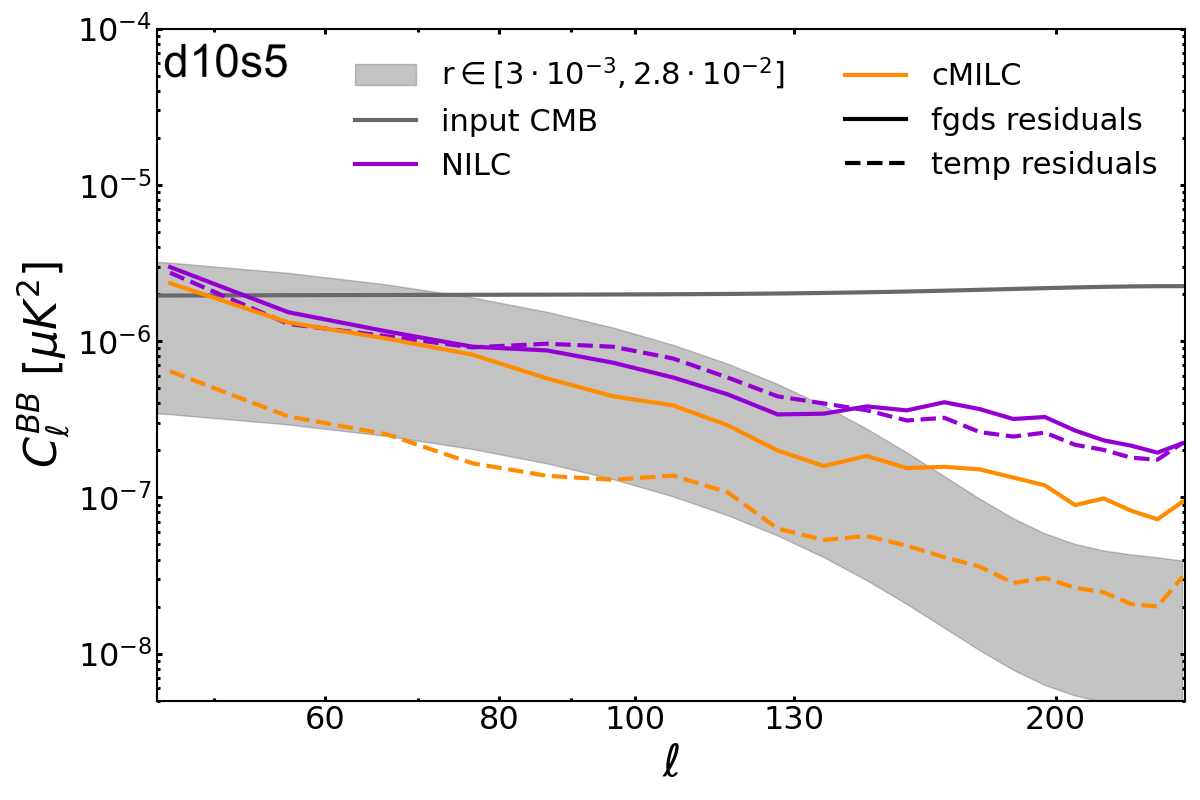}
\caption{$B$-mode angular power spectra of foreground residuals (solid lines) for NILC (violet) and cMILC (orange). Dashed lines instead report spectral templates of foreground residuals derived according to the procedure outlined in Section \ref{sec: margnalization}. All angular power spectra represent an average over 300 simulations. A uniform binning with $\Delta \ell=10$ is adopted. The average input CMB (with $r=0$) power spectrum obtained after applying the same masking to 300 simulated CMB-only maps, which contain lensing, is shown with the grey solid line. 
Results are reported for two foreground scenarios: \texttt{d1s1} (left panel) and \texttt{d10s5} (right panel), and assuming the baseline optimistic noise model. The grey shaded area indicates the amplitude range targeted by SO for the primordial tensor signal: $r \in [0.003, 0.028]$.} \label{fig: spectra template}
\end{figure}

Figure~\ref{fig: spectra template} shows the averaged power spectra of the foreground and noise residuals in the reconstructed CMB $B$-mode signal, obtained from the NILC and cMILC component-separation pipelines together with the corresponding residual spectral template. Results are reported for both the \texttt{d1s1} and \texttt{d10s5} scenarios. 
The spectral templates of the residual foreground contamination show good agreement with the shape of the actual foreground residuals. For both foreground scenarios, the agreement also extends to the amplitudes of the two angular power spectra in the NILC case, while a nearly constant mismatch is observed for cMILC. The observed mismatch arises from differences between the input foreground emission in the simulated data and the GNILC-reconstructed templates, which are further amplified when propagated through the component-separation pipeline using cMILC weights, more so than in the NILC. The mismatch of the low-frequency templates with respect to the input foreground signal (see Figure \ref{fig: templates}) does not have a significant impact on the recovered spectral template of the foreground residuals, since the corresponding channels receive much smaller weights in the component-separation process due to their low CMB sensitivity and dominant synchrotron emission.

\subsection{Results on parameter estimation} \label{sec: results with (m)}

\begin{figure}[tbp]
\centering
\includegraphics[width=.49\textwidth]{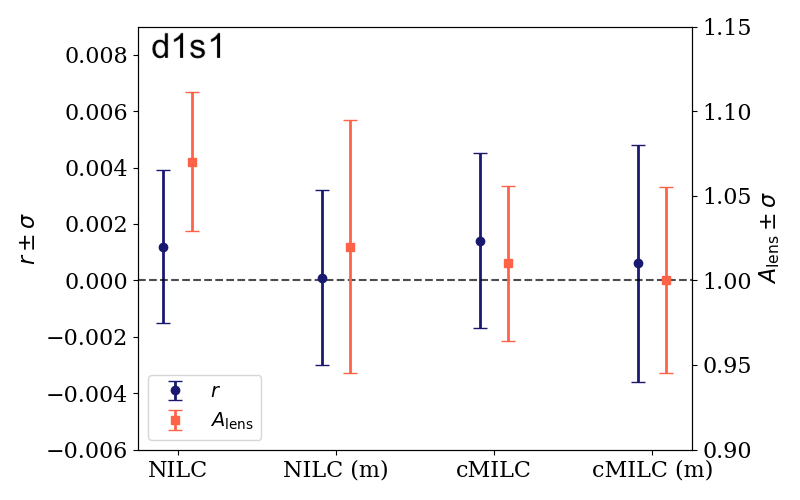 }
\includegraphics[width=.49\textwidth]{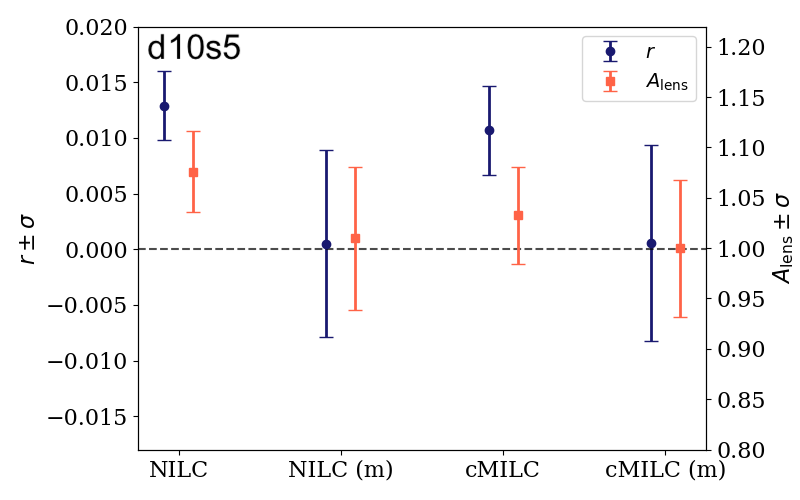}
\caption{Constraints from the $r$ and lensing amplitude $A_{\mathrm{lens}}$ posteriors obtained when sampling the cosmological likelihood in Equation \ref{eq: gaussian likelihood}. The cases where a template of foreground residuals is included in the spectral model are denoted with \texttt{(m)} in the labels. Results from the application of NILC and cMILC pipelines are presented. Error bars denote the $1\sigma$ credible intervals from the posterior distributions. Results are reported for foreground models: \texttt{d1s1} (left panel) and \texttt{d10s5} (right panel), and assuming the baseline optimistic noise model.}
 \label{fig: whiskerplot}
\end{figure}

We now report the constraints from the parameter estimation that we get in the \textit{marginalisation} case, where the spectral template of foreground residuals is included in the likelihood model through Equation~\ref{eq: model temp}. The corresponding results are denoted with the label \texttt{(m)}. Figure~\ref{fig: whiskerplot} shows the recovered best-fit values of the cosmological parameters, along with their corresponding $1\sigma$ confidence intervals, for all considered foreground models and component-separation pipelines, both with and without marginalisation over foreground residuals, and for the baseline-optimistic noise scenario. The corresponding quantitative results, reported in Table~\ref{tab: results}, show that incorporating the spectral template of foreground residuals into the likelihood model significantly reduces the residual bias in $r$: from $\mathcal{O}(10^{-3})$ for the \texttt{d1s1} case and $\mathcal{O}(10^{-2})$ for \texttt{d10s5} to undetectable levels ($\mathcal{O}(10^{-4})$), for both NILC and cMILC. At the same time, the marginalisation procedure yields unbiased estimates of $A_{\mathrm{lens}}$, therefore fully capturing the impact of component-separation foreground residuals. 

\begin{table}
\centering
\setlength{\tabcolsep}{6pt}
\renewcommand{\arraystretch}{1.2}
\begin{tabular}{l|ccc|ccc}
\hline
\textbf{Methods} 
& \multicolumn{3}{c|}{$(r \pm \sigma(r)) \cdot 10^{-3}$} 
& \multicolumn{3}{c}{$A_{\text{lens}}\pm\sigma(A_{\text{lens}})$} \\
& \multicolumn{2}{c}{Baseline} &Goal  & \multicolumn{2}{c}{Baseline} &Goal\\

& \textbf{d1s1} & \textbf{d10s5} & \textbf{d10s5} & \textbf{d1s1} & \textbf{d10s5} & \textbf{d10s5} \\
\hline
NILC       & $1.2 \pm 2.7$   & $12.9 \pm 3.1 $ & $13 \pm 2.7$& $1.07 \pm 0.04$  & $1.10 \pm 0.04$& $1.04\pm 0.03$\\
NILC (m)   & $0.1 \pm 3.1$ & $0.5 \pm 8.4$ &$-0.6\pm6.3$& $1.03 \pm 0.07$     & $1.01 \pm 0.07$& $1.01\pm0.03$ \\
cMILC      & $1.4 \pm 3.1$   & $10.7 \pm 4.0$ &$10.7\pm2.9$& $1.01 \pm 0.05$ & $1.03 \pm 0.05$&$1.03\pm0.03$ \\
cMILC (m)  & $0.6 \pm 4.2$   & $0.6 \pm 8.8$ &$0.7\pm7.2$& $1.00 \pm 0.05$ & $1.00 \pm 0.06$&$1.01\pm 0.04$ \\
\hline
\end{tabular}
\caption{Estimated best-fit values and $1\sigma$ uncertainty of the tensor-to-scalar ratio $r$ and the lensing amplitude $A_{\rm lens}$ for NILC and cMILC, with and without marginalisation (m) over foreground residuals for input $r=0$ and $A_\text{lens}=1$. Reported results refer to all considered combinations of foreground scenarios and instrumental noise assumptions.}
\label{tab: results}
\end{table}

This bias mitigation, obtained even in the more challenging \texttt{d10s5} scenario, is achieved at the cost of an increase in variance, as expected.  For the \texttt{d1s1} configuration and baseline-optimistic noise, the uncertainty on $r$ increases from $\sim 2.7$--$3.1 \times 10^{-3}$ to $\sim 3.1$--$4.2 \times 10^{-3}$, this is comparable to the degradation observed when analogous simulated data-sets are processed with similar extended pipelines \cite{Wolz_2024}, e.g. the cross-$C_{\ell}$ pipeline $+$ moments (Pipeline A) and the map-based parametric component separation $+$ marginalisation over foreground residuals (Pipeline C). This confirms that our template marginalisation captures a similar level of foreground uncertainty to that modelled in these extended analyses and that, with this addition, the blind component separation method can achieve similar performance as the other two pipelines developed for SO.
We also stress a methodological difference in how the foreground residual power spectrum template is obtained in our approach with respect to the one used for map-based parametric component separation in \cite{Wolz_2024}. In our approach, the templates are derived using GNILC and then propagated through the NILC and cMILC weights to capture the actual pipeline-specific residuals. In contrast, in Pipeline C of \cite{Wolz_2024}, implementing a parametric method, first the spectral parameters of dust and synchrotron are fitted on the multi-frequency maps, and then the sky components are reconstructed via a generalised least-squares inversion. To account for residual contamination in the clean CMB map after component separation, the likelihood model is augmented with a dust template derived from the reconstructed dust map, with its amplitude treated as a free nuisance parameter. While both methods remove the bias in $r$, the parametric approach depends on the accuracy of the foreground model, whereas our procedure directly targets pipeline-specific residuals, ensuring unbiased estimates of $r$ without any prior on foreground spectral behaviour.

For the \texttt{d10s5} case, despite the complete mitigation of foreground-induced biases, we obtain a significantly higher uncertainty on $r$ ($\sigma(r) \sim 8.4 \times 10^{-3}$), driven by the larger amplitude of foreground residuals that must be marginalised over, which contribute additional variance to the fit. For this case, we repeat the full analysis both component separation and cosmological inference using simulated SO-SATs–like data that include an enhanced noise model, the \emph{goal-optimistic} configuration, featuring a lower white-noise floor (see Section~\ref{sec: inputs} for details). The resulting constraints, reported in Table~\ref{tab: results}, show that the uncertainties on the cosmological parameters ($r$ and $A_{\mathrm{lens}}$) are significantly reduced, with $\sigma(r)$ decreasing to $6.3 \times 10^{-3}$ for NILC and $7.2 \times 10^{-3}$ for cMILC. In this case, our uncertainties on $r$ are larger than the ones reported in \cite{Wolz_2024} for the same noise scenario, while the bias is slightly smaller. We nevertheless note that the 
\texttt{d10s5} foreground model used in our work does not match perfectly the one considered in \cite{Wolz_2024}, as the PySM package has been updated to a new version, including modifications to both synchrotron and dust templates for this model; therefore, a direct comparison cannot be applied.

\begin{figure}[tbp]
\centering
\includegraphics[width=0.49\textwidth]{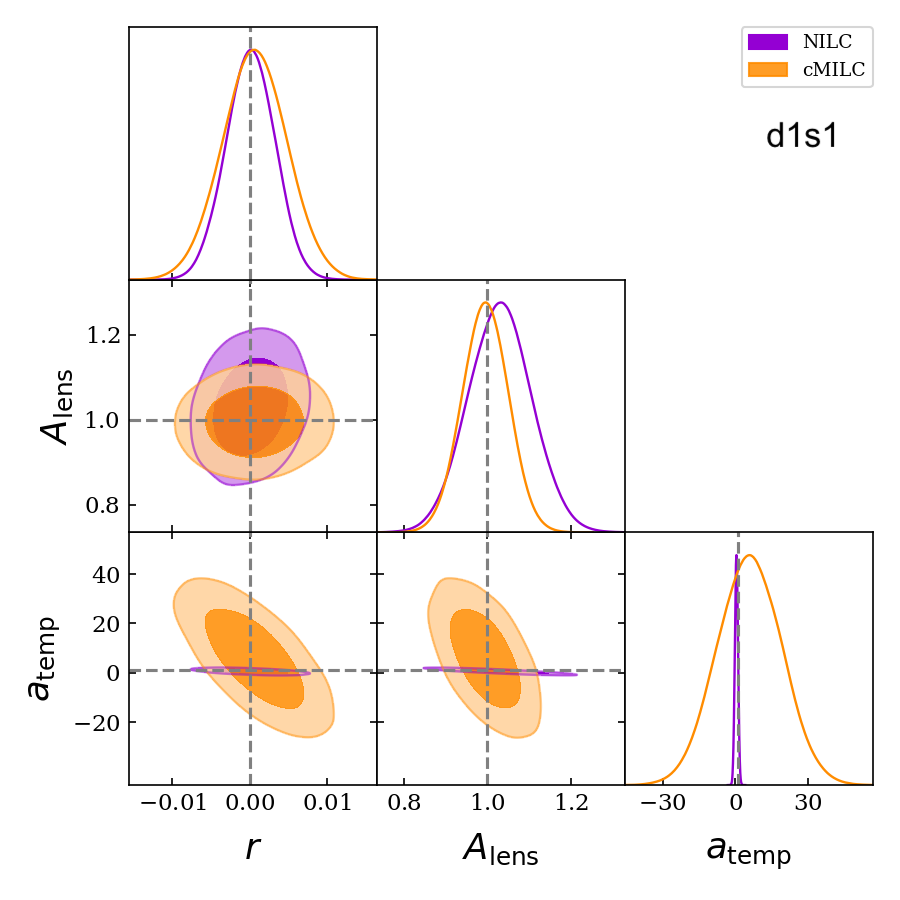}
\includegraphics[width=0.49\textwidth]{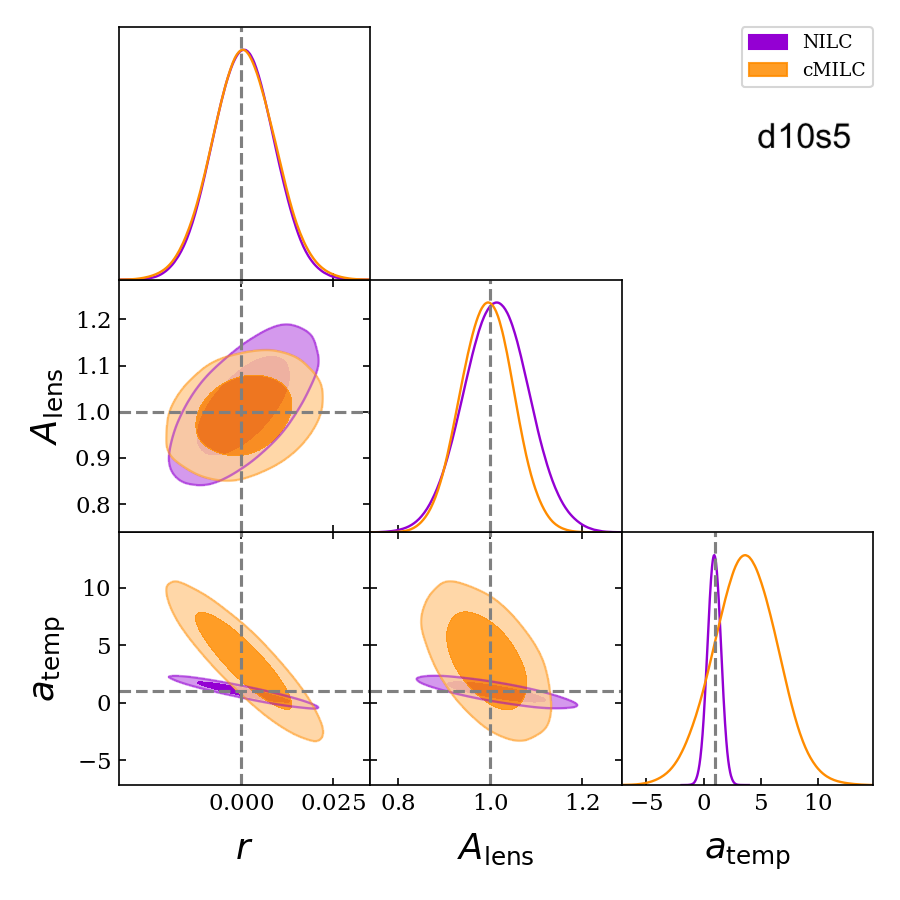}
\caption{Corner plots showing the posterior distributions for all the model parameters ($r$, $A_{\rm lens}$, and $a_{\rm temp}$) obtained from parameter inference using the mean reconstructed $B$-mode power spectrum averaged over simulations for NILC (violet) and cMILC (orange). Results are shown for the (\texttt{d1s1}, left) and (\texttt{d10s5}, right) foreground scenarios, and assuming the baseline optimistic noise model. Vertical dashed lines indicate the fiducial values $r=0$, $A_{\rm lens}=1.0$, and $a_{\rm temp}=1$.}
 \label{fig: cornor plots}
\end{figure}

The corner plots in Figure \ref{fig: cornor plots} report the joint posterior distributions of the three sampled parameters, $r$, $A_{\rm lens}$, and $a_{\rm temp}$, for both NILC and cMILC pipelines. The overall pattern of the posterior distribution is similar for the \texttt{d1s1} and \texttt{d10s5} foreground scenarios. 
We note that, since for NILC the residuals template closely matches the power amplitude of actual contamination, the posterior of $a_\textbf{temp}$ peaks at $\sim1$. For cMILC, instead given the difference in amplitude, the width of $a_\textbf{temp}$ posterior is significantly magnified, still allowing though to debias cosmological parameters inference.



For the same combinations of foreground and noise models considered in Table~\ref{tab: results}, we repeat the likelihood analysis by fixing $A_{\mathrm{lens}} = 1$ and jointly sampling only $r$ and $a_{\mathrm{temp}}$. This has the advantage of potentially reducing the overall uncertainty on the tensor-to-scalar
ratio, as the posterior distribution is sampled over a smaller parameter space. However, any residual systematic contamination of the data will then manifest itself as a bias exclusively in the
recovered value of $r$. The resulting confidence intervals for $r$, for all the considered component-separation pipelines, are reported in Table~\ref{tab: results A_lens=1}.
The most pronounced improvement in sensitivity to $r$ is obtained for the NILC application to the \texttt{d10s5} dataset, with $\sigma(r)$ improving from $8.4$ to $6.5$ when marginalisation is applied, and from $3.1$ to $2.6$ without it
 For both \texttt{d1s1} and \texttt{d10s5}, marginalisation over the spectral template continues to effectively mitigate foreground-induced biases, yielding unbiased estimates of $r$. 

\begin{table}
\centering
\begin{tabular}{l|ccc}
\hline
\textbf{Methods}
& \multicolumn{3}{c}{$(r \pm \sigma(r)) \cdot 10^{-3}$} \\
& \multicolumn{2}{c}{Baseline} &Goal \\
 & \textbf{d1s1} & \textbf{d10s5} & \textbf{d10s5}  \\
\hline
NILC       & $4.2 \pm 2.2$   & $17.0 \pm 2.6 $ & $14.4 \pm 2.3 $  \\
NILC (m)   & $-0.1 \pm 3.0$ & $-0.4 \pm 6.5$ & $-0.6 \pm 4.6$\\
cMILC      & $2.2 \pm 2.5$   & $10.7 \pm 4.0$ & $12.3 \pm 2.5$\\
cMILC (m)  & $0.7 \pm 4.2$   & $0.8 \pm 8.5$ & $0.5 \pm 7.4$ \\
\hline
\end{tabular}
\caption{Estimated best-fit values and $68\%$ confidence levels of the tensor-to-scalar ratio $r$ for NILC and cMILC, with and without marginalisation (m), with fixed $A_{\text{lens}}=1$ in the sampling of the cosmological likelihood.}
\label{tab: results A_lens=1}
\end{table}
\begin{table}
\centering
\setlength{\tabcolsep}{6pt}
\renewcommand{\arraystretch}{1.2}
\begin{tabular}{l|ccc|ccc}
\hline
\textbf{Methods} 
& \multicolumn{3}{c|}{$(r \pm \sigma(r)) \cdot 10^{-3}$} 
& \multicolumn{3}{c}{$A_{\text{lens}}\pm\sigma(A_{\text{lens}})$} \\
& \multicolumn{2}{c}{Baseline} &Goal  & \multicolumn{2}{c}{Baseline} &Goal\\

& \textbf{d1s1} & \textbf{d10s5} & \textbf{d10s5} & \textbf{d1s1} & \textbf{d10s5} & \textbf{d10s5} \\
\hline
NILC       & $0.9\pm 2.3$   & $12.7 \pm 2.6 $ & $11.2 \pm 1.9$& $0.64 \pm 0.03$  & $0.60 \pm 0.03$& $0.54\pm 0.02$\\
NILC (m)   & $0.3 \pm 2.6$ & $0.2 \pm 6.9$ &$-0.5\pm4.9$& $0.50 \pm 0.06$ & $0.51 \pm 0.06$& $0.51\pm0.02$ \\
cMILC      & $1.1 \pm 2.7$   & $10.4 \pm 3.3$ &$10.2\pm2.2$& $0.51 \pm 0.04$ & $0.53 \pm 0.04$&$0.53\pm0.02$ \\
cMILC (m)  & $0.8 \pm 3.3$   & $0.6 \pm 7.2$ &$0.8\pm5.5$& $0.50 \pm 0.04$ & $0.51 \pm 0.05$&$0.50\pm 0.03$ \\
\hline
\end{tabular}
\caption{Estimated best-fit values and $1\sigma$ uncertainty of the tensor-to-scalar ratio $r$ and the lensing amplitude $A_{\rm lens}$ for NILC and cMILC, with and without marginalisation (m) over foreground residuals for input $r=0$ and $A_\text{lens}=0.5$. Reported results refer to all considered combinations of foreground scenarios and instrumental noise assumptions.}
\label{tab: results with A_lens=0.5}
\end{table}
For the same combinations of foreground and noise models considered in Table \ref{tab: results}, we repeat the likelihood analysis but implement a 50\% delensing at the spectrum level in the output maps. Therefore, in this case, the fiducial $A_{\rm lens}$ value amounts to $0.5$. In practice, for each simulation this is achieved by subtracting $(1-\sqrt{A_{\rm lens}}) \cdot \textrm{B}^{\rm lens}(\hat{n})$ from the reconstructed CMB map, where $\textrm{B}^{\rm lens}(\hat{n})$ is the input lensing $B$-moded signal. The corresponding results are reported in Table \ref{tab: results with A_lens=0.5} and feature the expected reduction in the recovered lensing amplitude, with uncertainties for both cosmological parameters significantly reduced compared to the $A_\text{lens}=1$ case due to lower lensing $B$-mode variance. Importantly, marginalising over the foreground spectral template remains effective, producing unbiased estimates for both $r$ and $A_\text{lens}$. Our results for the 50\%  delensing in case of NILC for \texttt{d1s1} model are consistent with \cite{Wolz_2024}(Table 5). Notably, the uncertainty on $r$ for the more complex foreground scenario (\texttt{d10s5}) decreases from $\sigma(r)=8.4$ to $\sigma(r)=6.9$ (for baseline optimistic noise model) and from $\sigma(r)=6.3$ to $\sigma(r)=4.9$ (for goal optimistic noise scenario) when applying NILC component separation and marginalisation.

\begin{figure}
    \centering
    \includegraphics[width=0.65\linewidth]{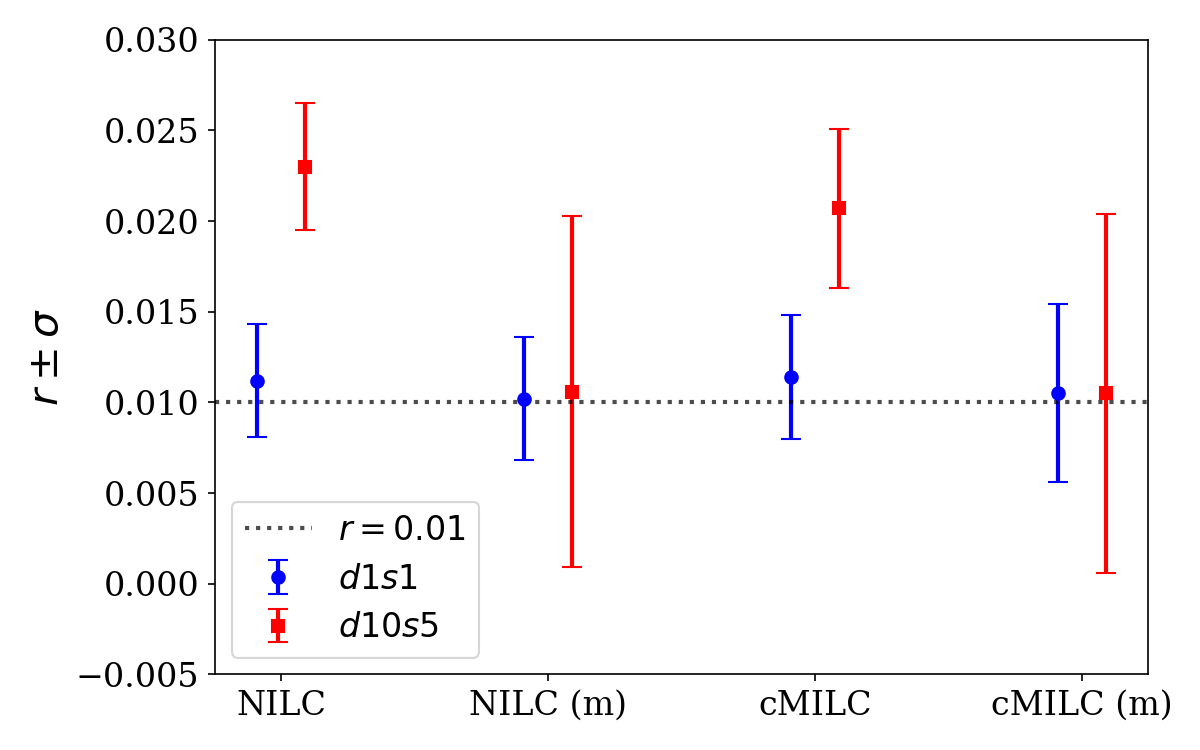}
    \caption{Constraints from the $r$ posterior obtained when sampling the cosmological likelihood in Equation \ref{eq: gaussian likelihood}, with input model containing primordial $B$-modes with $r=0.01$. The label \texttt{(m)} denotes marginalisation over foreground template residuals in the spectral model. Results from the application of NILC and cMILC pipelines are presented. Error bars denote the $1\sigma$ credible intervals from the posterior distributions. Results are reported for foreground models: \texttt{d1s1} (blue) and \texttt{d10s5} (red), and assuming the baseline optimistic noise model.}
    \label{fig: tensor posterior}
\end{figure}

As a final step, we repeat the same inference analysis, this time including a primordial tensor component in the input CMB $B$-mode signal, with a tensor-to-scalar ratio of $r_{\mathrm{in}} = 0.01$, and with no delensing ($A_{\rm lens}=1)$. Since the residuals from component separation are not expected to depend strongly on the overall amplitude of the tensor modes, given that the CMB contributes as a constant signal in the minimum-variance combination and therefore only adds a constant term to the covariance, neglecting chance correlations, we did not rerun the full component-separation pipeline. Instead, we manually added simulated tensor-induced $B$ modes to the recovered CMB maps and repeated the computation of the angular power spectra, covariance, and likelihood. In this case, we retain the baseline likelihood model, also sampling $A_{\mathrm{lens}}$. The corresponding credible intervals are shown in Figure~\ref{fig: tensor posterior} (for the tensor-to-scalar ratio) and reported in Table~\ref{tab: results with r=0.01}. We find that even for a non-zero input $r$, the inferred best-fit value remains biased by foreground residuals. However, when marginalisation is implemented, the input value is accurately recovered, confirming both the effectiveness and robustness of the adopted approach. As expected, the $1\sigma$ uncertainties on $r$ increase due to the additional cosmic variance introduced by the tensor-mode contribution to the observed $B$-mode power spectrum.

\begin{table}
\centering
\begin{tabular}{l|cc|cc}
\hline
\textbf{Methods}
& \multicolumn{2}{c|}{$(r \pm \sigma(r)) \cdot 10^{-3}$}
& \multicolumn{2}{c}{$A_{\text{lens}} \pm \sigma(A_{\text{lens}})$} \\
 & \textbf{d1s1} & \textbf{d10s5} & \textbf{d1s1} & \textbf{d10s5} \\
\hline
NILC   & $11.2 \pm 3.1$ & $ 22.9 \pm 3.5$ & $1.08 \pm 0.04$ & $ 1.10\pm0.04 $ \\
NILC (m)   & $10.2 \pm 3.4$ & $10.5 \pm  9.7$ & $1.03 \pm 0.08$ & $1.02 \pm 0.08$ \\
cMILC  & $11.4 \pm 3.4$   & $20.8 \pm 4.4$ & $1.01 \pm 0.05$ & $1.03\pm 0.05$ \\
cMILC (m)  & $10.6 \pm 4.9$   & $10.5 \pm 9.9$ & $1.00 \pm 0.06$ & $1.00 \pm 0.06$ \\
\hline
\end{tabular}
\caption{Estimated best-fit values and $1\sigma$ uncertainty of the tensor-to-scalar ratio $r$ and the lensing amplitude $A_{\rm lens}$ for NILC and cMILC, with and without marginalisation (m) over foreground residuals, with the input model including primordial $B$-modes with $r=0.01$.}
\label{tab: results with r=0.01}
\end{table}
Overall, we conclude that the inclusion of foreground marginalisation enables unbiased estimates of cosmological parameters from realistic simulated data of ongoing ground-based CMB experiments with sky and frequency coverage comparable to that proposed for the SO-SATs. This conclusion holds across different assumed foreground models and input values of the tensor-to-scalar ratio $r$.

\section{Conclusions} \label{sec: conclusion}

Several ongoing or planned CMB experiments are pursuing the ambitious goal of either detecting a primordial $B$-mode signal or significantly tightening the current upper limits on its amplitude. Achieving this objective requires the development of highly effective component-separation techniques capable of removing the bulk of the dominant Galactic polarised emission. This challenge is already faced by current ground-based experiments, such as BICEP/Keck~\cite{BK2021} and SO~\cite{Ade_2019}.
For instance, a recent forecast of the expected sensitivity of different component-separation pipelines to the tensor-to-scalar ratio $r$ has been presented in Ref.~\cite{Wolz_2024} for the SO-SATs. That analysis shows that component-separation pipelines based on parametric methods applied to multifrequency power spectra, or on map-based simulations, can achieve unbiased constraints on $r$ when appropriate extensions of the pipelines are considered. In contrast, when adopting a model-independent approach specifically the NILC pipeline residual foreground leakage in the reconstructed $B$-mode angular power spectrum can lead to biased estimates of cosmological parameters, particularly in the presence of more realistic and complex foreground scenarios.

Therefore, in this paper, we explore extensions of the baseline NILC pipeline aimed at mitigating the impact of foreground residuals on cosmological inference from the reconstructed CMB $B$-mode signal expected to be recovered from data of ongoing ground-based CMB experiments. Specifically, we consider two not mutually exclusive approaches: (i) the deprojection of statistical moments of foreground emission through the constrained moment ILC (cMILC) technique \cite{Remazeilles_2020} (see Section~\ref{sec:cMILC}); and (ii) the marginalisation over a spectral template of foreground residuals within the cosmological likelihood framework \cite{Carones_marg} (see Section~\ref{sec: marg l(r)}). Given that observations with the SO-SATs just began in 2024 and that a recent forecast including map-based component-separation pipelines is available for comparison~\cite{Wolz_2024}, we validate these extensions using realistic simulations of SO-SAT observations.

We apply these extensions together with the standard NILC pipeline to 300 different realistic simulations, analogous to those considered in \cite{Wolz_2024}, making use of the public Python package \texttt{BROOM}. We consider two different foreground models, featuring increasing complexity: the \texttt{PySM} \texttt{d1s1} and \texttt{d10s5} (see Section \ref{sec: inputs}). Averaged angular power spectra of residuals obtained from component separation are shown in Figure~\ref{fig: whiskerplot}, while summary statistics of the posteriors of the fitted cosmological parameters, tensor-to-scalar ratio $r$ and lensing amplitude $A_{\text{lens}}$, are reported in Table~\ref{tab: results}. The main results can be summarised as follows:

\begin{itemize}
    \item The cMILC method provides an improvement over standard NILC in reducing foreground residuals, especially when marginalisation over the foreground spectral template is not performed, highlighting the advantage of deprojecting leading foreground moments to suppress bias, although it comes with a penalty of higher variance.
    \item The inclusion of spectral foreground templates in the likelihood model effectively removes these biases for NILC and cMILC, even in the challenging case of \texttt{d10s5}, providing unbiased lensing amplitude and tensor-to-scalar ratio estimates.

\end{itemize}

The above results demonstrate that the considered extensions to the standard NILC pipeline provide a robust framework for unbiased estimation of $r$ in the presence of realistic Galactic foregrounds and in the context of ground-based CMB experiments, such as SO. Possible further improvements are foreseen by applying the NILC minimum variance solution to optimised sky domains \cite{carones_mcnilc} and by including additional external data-sets in the component separation step, able to increase the leverage arm on foreground spectral scaling. We defer these analyses for future work.




\acknowledgments
This paper and related research have been conducted during and with the support of the Italian national inter-university PhD program in Space Science and Technology, CUP master: I53D24000060005 and CUP SISSA: G93C24000890006. We acknowledge partial support by the Italian Space Agency LiteBIRD Project (ASI Grants No. 2020-9-HH.0 and 2016-24-H.1-2018), and LiteBIRD Initiative of the National Institute for Nuclear Physics, and Project SPACE-IT-UP by the Italian Space Agency and Ministry of University and Research, Contract Number 2024-5-E.0 and the RadioForegroundsPlus Project HORIZON-CL4-2023-SPACE-01, GA 101135036, and the CMB-Inflate project funded by the European Union’s Horizon 2020 Research and Innovation Staff Exchange under the Marie Skłodowska-Curie grant agreement No 101007633. This work received support from computational resources provided by the National Energy Research Scientific Computing Centre (NERSC), which is managed by the Lawrence Berkeley National Laboratory for the U.S. Department of Energy. 
This work is not an official Simons Observatory paper.





\bibliographystyle{JHEP}
\bibliography{sample}

\end{document}